\documentclass[aps,prb,reprint,10pt,superscriptaddress,floatfix,nofootinbib]{revtex4-2}

\usepackage{amsmath}
\usepackage[T1]{fontenc}
\usepackage[utf8]{inputenc}
\usepackage{lmodern}
\usepackage{amssymb}
\usepackage{natbib}
\usepackage{graphicx}
\usepackage{hyperref}
\usepackage{epstopdf}
\usepackage{placeins}
\usepackage{orcidlink}
\usepackage{bm}

\graphicspath{{figures/}}

\hypersetup{
	colorlinks=true,
	linkcolor=blue,
	citecolor=blue,
	urlcolor=blue
}

\newcommand{\Om}{\Omega}
\newcommand{\etaT}{\eta}
\newcommand{\Ceff}{C_{\rm eff}}
\newcommand{\EC}{E_C}
\newcommand{\EJ}{E_J}
\newcommand{\Dphi}{\Delta_\phi}
\newcommand{\Dz}{\Delta_z}

\begin{document}
	
	\title{Synthetic twist geometry in mesoscopic quantum circuits:\\ geometric capacitance and chiral mode splitting}
	
	\author{Edilberto O. Silva\orcidlink{0000-0002-0297-5747}}
	\email[Edilberto O. Silva - ]{edilberto.silva@ufma.br}
	\affiliation{
		Programa de P\'os-Gradua\c c\~ao em F\'{\i}sica \&
		Coordena\c c\~ao do Curso de F\'{\i}sica -- Bacharelado,
		Universidade Federal do Maranh\~{a}o,
		65085-580 S\~{a}o Lu\'{\i}s, Maranh\~{a}o, Brazil}
	
	\begin{abstract}
		We propose a route for encoding a helicoidally twisted synthetic geometry in mesoscopic quantum circuits.  A torsionless spatial metric with twist parameter $\Om$ defines an electrostatic capacitance kernel whose angular--axial sector contains the chiral coupling $-2\Om mk$.  We show that this kernel can be implemented by a finite, quantizable circuit graph built only from positive two-node capacitors: diagonal bridges generate the cross term, and the dimensionless twist is the ratio of diagonal to total angular capacitance.  The resulting capacitance matrix enters the Hamiltonian through $C^{-1}(\etaT)$ and splits counter-rotating synthetic modes.  Exact finite-graph diagonalization confirms the analytic spectrum and shows that closed longitudinal circulation is required; with open boundaries the bridge phase is gauge removable and the doublets remain degenerate.  Simulated spectroscopy resolves the chiral fan, while disorder simulations indicate that calibrated percent-level reactive disorder does not obscure the deterministic splitting in the parameter range considered.  Twist textures produce interface modes, Josephson nonlinearities make the Kerr sector chirality dependent, and the same doublet provides a detuning knob for non-Hermitian exceptional-point control.  For representative circuit-QED parameters the splitting is in the tens-of-MHz range, well above typical resonator linewidths.
	\end{abstract}
	
	\maketitle
	
	\section{Introduction}
	\label{sec:introduction}
	
	Geometry is a long-standing organizing principle in condensed matter.  In the continuum theory of crystal defects, elastic media containing dislocations and disclinations are described by effective non-Euclidean metrics, with screw dislocations generating mixed angular--axial terms of exactly the type studied here \cite{Katanaev1992}.  More recently, synthetic dimensions, artificial gauge structures, and engineered connectivity have made it possible to emulate such effective geometries in tunable platforms, where the metric becomes a design parameter rather than a property of a real material \cite{Celi2014,Yuan2018,OzawaPrice2019,Ozawa2019,Lustig2019,ArguelloLuengo2024}.
	
	Lattice electrical circuits are a particularly flexible member of this family.  Topolectrical and lumped-element networks realize engineered band structures directly at the level of the admittance matrix \cite{Ningyuan2015,Lee2018}, and networks of coplanar-waveguide resonators encode non-Euclidean structures, including hyperbolic lattices and curved-space tight-binding models \cite{Kollar2019,Boettcher2020}.  Superconducting circuits add the mesoscopic, quantizable limit: their variables are first described by classical fluxes and charges and are then canonically quantized, so that classical electromagnetic response functions translate directly into engineered Hamiltonians at the single- or few-photon level \cite{Wallraff2004,Nigg2012,Vool2017,Koch2007,Houck2012,Blais2021}.  These advances motivate the question addressed here: can a continuous helicoidal metric be converted into a mesoscopic circuit Hamiltonian whose spectrum is controlled by a geometric twist?
	
	We develop a minimal theory of \emph{twist quantum circuits}.  The helicoidal metric is used as a target synthetic geometry encoded first in the electrostatic energy, then in a discrete capacitance matrix, and finally in the quantum Hamiltonian.  The classical calculation is therefore a bridge, not the endpoint: it identifies the capacitance kernel that a circuit graph must reproduce.
	
	The target geometry is the torsionless helicoidal spatial line element \cite{SilvaNettoFurtado2008,BakkeMoraes2012PLA,Silva2026AnnPhys,Silva2026OQE}
	\begin{equation}
		\label{eq:line_element}
		dl^2=dr^2+r^2d\phi^2+
		\left(dz+\Om r^2d\phi\right)^2 .
	\end{equation}
	Here $\Om$ is a metric twist, not a Cartan torsion.  Only the spatial metric enters the construction, so no relativistic convention is required; all electrostatic and circuit quantities below are expressed in SI units.  Since $z$ and $r$ have dimensions of length while $\phi$ is dimensionless, $\Om$ has dimension of inverse length.  The natural dimensionless control parameter is
	\begin{equation}
		\label{eq:eta_def}
		\etaT=\Om R,
	\end{equation}
	where $R$ is the transverse size of the effective circuit region.  The central signature is the angular--axial term $-2\Om mk$, which separates modes proportional to $e^{i(m\phi+kz)}$ from their counter-rotating partners.  The main purpose of the discrete construction below is to show how this term can arise from an explicitly positive circuit capacitance matrix.  In practical terms, the synthetic twist acts as a layout- or coupler-controlled geometric detuning knob: it separates helical modes without magnetic flux, provides phase-matched spectroscopic selectivity, and can be combined with Josephson nonlinearities or engineered loss channels to tune Kerr interactions and exceptional-point conditions.
	
	The paper is organized as follows.  Section~\ref{sec:geometry} summarizes the helicoidal metric and its angular--axial mixing.  Section~\ref{sec:classical_capacitance} computes the classical capacitance kernel and identifies which electrode configurations respond to the twist.  Section~\ref{sec:circuit_reduction} promotes the kernel to a mesoscopic Hamiltonian.  Section~\ref{sec:discrete_circuit} contains the central construction: the discrete twist kernel, its realization as an explicit two-node capacitor netlist, and the quantitative dictionary back to the continuum.  Section~\ref{sec:chiral_modes} presents the chiral spectra and their observability: sheared iso-frequency contours, exact diagonalization of a finite graph, simulated chiral spectroscopy, disorder robustness, and SI estimates.  Section~\ref{sec:domain_walls} studies twist textures---domain walls and heterojunctions---and their interface modes.  Section~\ref{sec:kerr} introduces Josephson nonlinearity and derives twist-controlled Kerr interactions.  Section~\ref{sec:nonhermitian} develops the non-Hermitian two-mode sector, and Secs.~\ref{sec:discussion} and \ref{sec:conclusions} close with outlook and conclusions.
	
	\section{Helicoidal metric and angular--axial mixing}
	\label{sec:geometry}
	
	Starting from Eq.~\eqref{eq:line_element}, the inverse metric components needed below are
	\begin{align}
		\label{eq:inverse_metric}
		g^{rr}&=1, &
		g^{\phi\phi}&=\frac{1}{r^2}, &
		g^{\phi z}&=-\Om, &
		g^{zz}&=1+\Om^2r^2 ,
	\end{align}
	with volume element $\sqrt{g}=r$.  With the sign conventions
	$R^{\rho}{}_{\sigma\mu\nu}=\partial_\mu\Gamma^{\rho}_{\nu\sigma}
	-\partial_\nu\Gamma^{\rho}_{\mu\sigma}
	+\Gamma^{\rho}_{\mu\lambda}\Gamma^{\lambda}_{\nu\sigma}
	-\Gamma^{\rho}_{\nu\lambda}\Gamma^{\lambda}_{\mu\sigma}$,
	$R_{\sigma\nu}=R^{\rho}{}_{\sigma\rho\nu}$, and
	$\mathcal{R}=g^{\sigma\nu}R_{\sigma\nu}$, for which a sphere has
	$\mathcal{R}>0$, direct computation gives the scalar curvature
	\begin{equation}
		\label{eq:scalar_curvature}
		\mathcal{R}=-2\Om^2 .
	\end{equation}
	Thus the twist is a metric deformation, although the cylindrical volume element remains unchanged.  The response comes from the inverse metric, especially $g^{\phi z}$ and $g^{zz}$.
	
	For a mode factor $\Psi\sim f(r)e^{i(m\phi+kz)}$, the angular--axial contribution is
	\begin{align}
		\label{eq:q_contraction}
		g^{ij}q_iq_j
		&=\frac{m^2}{r^2}-2\Om mk+
		\left(1+\Om^2r^2\right)k^2 \\
		&=k^2+\left(\frac{m}{r}-\Om kr\right)^2 .
	\end{align}
	The linear term is odd under $m\to -m$ or $k\to -k$ and is the source of chiral splitting in the synthetic circuit modes.
	
	\section{Classical capacitance kernel}
	\label{sec:classical_capacitance}
	
	For a dielectric permittivity $\epsilon$, electrostatics in the spatial metric is governed by
	\begin{equation}
		\label{eq:electrostatic_energy}
		U_E[V]=\frac{\epsilon}{2}\int d^3x\sqrt{g}\,
		g^{ij}\partial_iV\partial_jV .
	\end{equation}
	Varying $V$ gives
	\begin{equation}
		\label{eq:laplace_general}
		\frac{1}{\sqrt{g}}\partial_i
		\left(\sqrt{g}\,g^{ij}\partial_jV\right)=0 .
	\end{equation}
	For prescribed electrode potentials, the effective capacitance is defined by $U_E=\frac{1}{2}\Ceff(\Om)V_0^2$.  The following benchmarks identify which coordinate configurations probe the twist.
	
	\subsection{Radial benchmark}
	\label{subsec:radial}
	
	For coaxial cylindrical electrodes at $r=a$ and $r=b$, with length $L$, impose $V(a)=V_0$ and $V(b)=0$.  A radial solution $V=V(r)$ sees only $g^{rr}=1$.  Since $\sqrt{g}=r$, the Laplace problem is the usual cylindrical one, with solution $V(r)=V_0\ln(b/r)/\ln(b/a)$ and capacitance
	\begin{equation}
		\label{eq:radial_cap}
		C_r(\Om)=\frac{2\pi\epsilon L}{\ln(b/a)} .
	\end{equation}
	This null result shows that the twist response is not produced by the measure alone.
	
	\subsection{Longitudinal coordinate capacitor}
	\label{subsec:longitudinal}
	
	Now take coordinate plates at $z=0$ and $z=d$ with transverse radius $R$.  This should be read as a synthetic or modal capacitor aligned with the coordinate $z$, not as a claim about literal Euclidean plates embedded in a curved laboratory.  The potential $V(z)=V_0(1-z/d)$ solves Eq.~\eqref{eq:laplace_general}.  Because $\partial_zV=-V_0/d$, the integrand of Eq.~\eqref{eq:electrostatic_energy} is $\epsilon g^{zz}V_0^2/2d^2$, and integrating $g^{zz}=1+\Om^2r^2$ over the cylinder gives
	\begin{equation}
		\label{eq:long_cap}
		C_z(\Om)=C_z(0)\left(1+\frac{\etaT^2}{2}\right),
		\qquad
		C_z(0)=\frac{\epsilon\pi R^2}{d} .
	\end{equation}
	The twist enhances the longitudinal capacitance and lowers the charging energy of a mesoscopic mode dominated by this capacitance.
	
	\subsection{Angular--axial modal capacitance}
	\label{subsec:modal_capacitance}
	
	For a voltage mode
	\begin{equation}
		\label{eq:modal_voltage}
		V_{mk}(r,\phi,z,t)=\mathcal{V}_{mk}(t)
		f_{mk}(r)e^{i(m\phi+kz)}+{\rm c.c.},
	\end{equation}
	the electrostatic functional contains the kernel
	\begin{align}
		\label{eq:modal_kernel}
		\mathcal{K}_{mk}(\Om)=
		\int_0^R r\,dr\,
		\Bigg[&\left|\frac{df_{mk}}{dr}\right|^2
		+\bigg(\frac{m^2}{r^2}-2\Om mk \\
		&+\left(1+\Om^2r^2\right)k^2\bigg)|f_{mk}|^2\Bigg] .
	\end{align}
	Completing the square as in Eq.~\eqref{eq:q_contraction}, the twist enters through $(m/r-\Om kr)^2$.  Unlike the longitudinal benchmark, this modal kernel contains a term linear in $\Om$.  It is the most distinctive geometric feature of the proposal.
	
	\section{From capacitance kernel to mesoscopic Hamiltonian}
	\label{sec:circuit_reduction}
	
\begin{figure*}[t]
		\centering
		\includegraphics[width=0.88\textwidth]{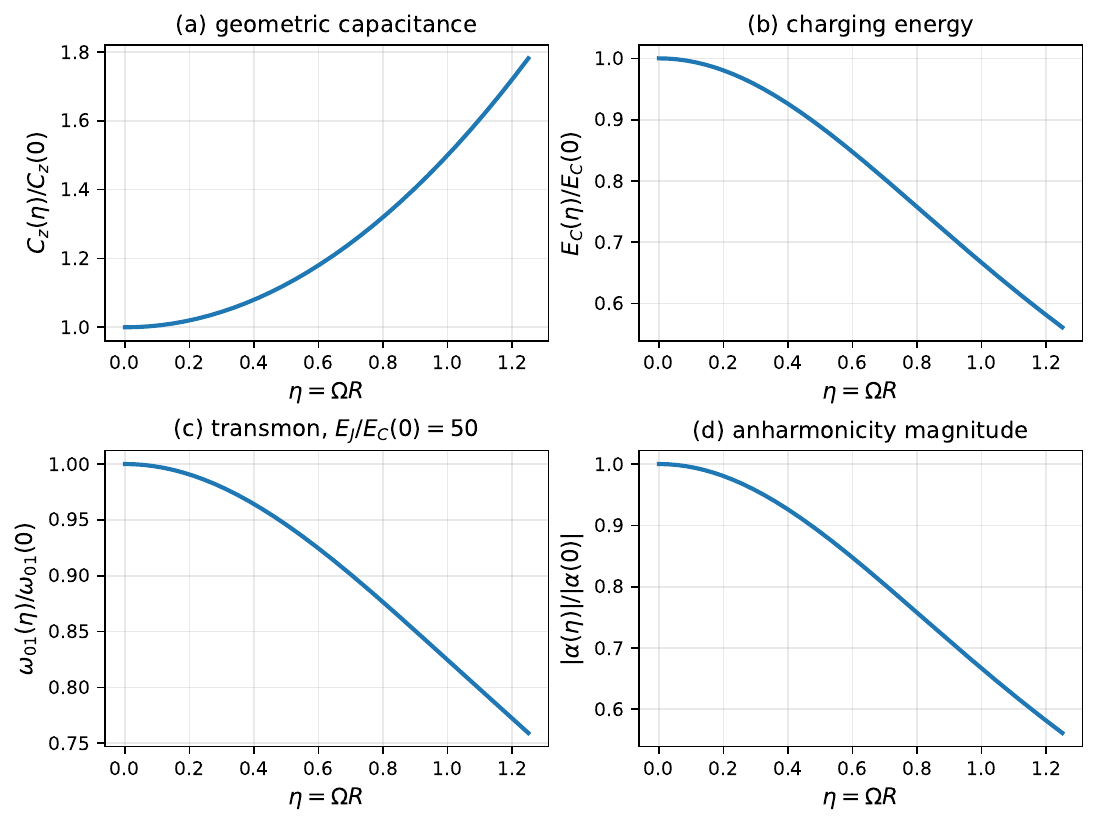}
		\caption{Illustrative classical-to-quantum consequence of the longitudinal geometric capacitance, assuming the shunt capacitance is dominated by the twist-dependent kernel.  The twist enhances $C_z$, reduces $E_C$, redshifts the transmon transition, and reduces the magnitude of the anharmonicity.  The transmon panel uses $E_J/E_C(0)=50$ as a representative value; the plotted ratios are dimensionless.}
		\label{fig:geometric_capacitance}
	\end{figure*}

	Project the voltage onto electrode or mode functions $f_a(\mathbf{x})$,
	\begin{equation}
		\label{eq:potential_expansion}
		V(\mathbf{x},t)=\sum_a V_a(t)f_a(\mathbf{x}) .
	\end{equation}
	Then Eq.~\eqref{eq:electrostatic_energy} becomes
	\begin{equation}
		\label{eq:cap_matrix_energy}
		U_E=\frac{1}{2}\sum_{a,b}C_{ab}(\Om)V_aV_b,
	\end{equation}
	where
	\begin{equation}
		\label{eq:cap_matrix}
		C_{ab}(\Om)=\epsilon\int d^3x\sqrt{g}\,
		g^{ij}\partial_if_a\partial_jf_b .
	\end{equation}
	With node fluxes $\Phi_a$ and $V_a=\dot\Phi_a$, the circuit Lagrangian is
	\begin{equation}
		\label{eq:circuit_lagrangian}
		\mathcal{L}=\frac{1}{2}\dot{\bm\Phi}^{T}C(\Om)
		\dot{\bm\Phi}-U(\bm\Phi).
	\end{equation}
	Legendre transformation and canonical quantization, with $[\hat\Phi_a,\hat Q_b]=i\hbar\delta_{ab}$, then give
	\begin{equation}
		\label{eq:quantum_hamiltonian_general}
		\hat H(\Om)=\frac{1}{2}\hat{\bm Q}^{T}C^{-1}(\Om)
		\hat{\bm Q}+U(\hat{\bm\Phi}) .
	\end{equation}
	For a single LC mode this gives
	\begin{equation}
		\label{eq:omega_lc}
		\omega_c(\etaT)=\omega_c(0)
		\left(1+\frac{\etaT^2}{2}\right)^{-1/2}.
	\end{equation}
	For a Josephson circuit, the same dependence enters through $\EC(\Om)=e^2/[2C_\Sigma(\Om)]$ in the standard Hamiltonian $4\EC(\hat n-n_g)^2-\EJ\cos\hat\varphi$.  Representative values $C_\Sigma\simeq65\,\mathrm{fF}$ give $\EC/h\simeq0.30\,\mathrm{GHz}$, the usual transmon scale \cite{Koch2007}.  If the dominant twist-dependent part of $C_\Sigma$ is the longitudinal geometric capacitance, then in the transmon regime
	\begin{align}
		\label{eq:EC_eta}
		\EC(\etaT)&=\EC(0)\left(1+\frac{\etaT^2}{2}\right)^{-1},\\
		\label{eq:omega01}
		\hbar\omega_{01}(\etaT)&\simeq
		\sqrt{8\EJ\EC(\etaT)}-\EC(\etaT),\\
		\label{eq:anharmonicity}
		\alpha(\etaT)&\simeq-\frac{\EC(\etaT)}{\hbar} .
	\end{align}
	This single-mode reduction is illustrative: it assumes that the total shunt capacitance of the device is dominated by the twist-dependent kernel.  The multimode graph constructed in Sec.~\ref{sec:discrete_circuit} does not rely on this assumption, since there the twist enters every eigenvalue of the capacitance matrix directly.  Figure~\ref{fig:geometric_capacitance} illustrates the basic spectral consequences of Eqs.~\eqref{eq:long_cap}--\eqref{eq:anharmonicity}.
	
	\section{Discrete circuit realization of the twist kernel}
	\label{sec:discrete_circuit}
	
	\subsection{Minimal twist kernel}
	\label{subsec:minimal_kernel}
	
	We now make the synthetic-circuit step explicit.  Consider a finite graph of node fluxes $\Phi_{j,n}(t)$, where $j=0,\ldots,N_\phi-1$ labels an angular synthetic coordinate and $n=0,\ldots,N_z-1$ labels a longitudinal coordinate.  The angular direction is periodic.  Define forward differences
	\begin{align}
		\label{eq:discrete_differences}
		\Dphi \Phi_{j,n}&=\Phi_{j+1,n}-\Phi_{j,n},\\
		\Dz \Phi_{j,n}&=\Phi_{j,n+1}-\Phi_{j,n}.
	\end{align}
	A minimal positive capacitive kinetic energy that realizes the twist kernel is
	\begin{align}
		\label{eq:discrete_TC_square}
		T_C=&\frac{1}{2}\sum_{j,n}\Big[
		C_g\dot\Phi_{j,n}^2+C_z\left(\Dz\dot\Phi_{j,n}\right)^2
		\nonumber\\
		&\hspace{1.1cm}+C_\phi\left(\Dphi\dot\Phi_{j,n}
		-\etaT\Dz\dot\Phi_{j,n}\right)^2\Big] .
	\end{align}
	The last term is the circuit analogue of the metric square
	$(m/r-\Om kr)^2$: its expansion contains the cross-capacitive contribution $-\etaT C_\phi\sum_{j,n}(\Dphi\dot\Phi_{j,n})(\Dz\dot\Phi_{j,n})$, the finite-graph counterpart of the continuum term $-2\Om mk$.  Since Eq.~\eqref{eq:discrete_TC_square} is a sum of squares with $C_g,C_z,C_\phi>0$, the capacitance matrix is positive by construction.
	
	As written, however, the last term of Eq.~\eqref{eq:discrete_TC_square} involves a weighted voltage difference among \emph{three} nodes, whereas a physical capacitor stores energy in the voltage difference between \emph{two} nodes.  After deriving the target spectrum, Sec.~\ref{subsec:netlist} shows that the same kernel is realized, at long wavelengths, by a network built exclusively from two-node capacitors.
	
	\subsection{Fourier spectrum and chiral splitting}
	\label{subsec:fourier_spectrum}
	
	For periodic boundary conditions, use the Fourier expansion
	\begin{equation}
		\label{eq:discrete_fourier}
		\Phi_{j,n}=\frac{1}{\sqrt{N_\phi N_z}}
		\sum_{p,q}\Phi_{p,q}e^{i(pj+qn)},
	\end{equation}
	where $p=2\pi m/N_\phi$ and $q=2\pi \ell/N_z$.  The capacitance eigenvalue is
	\begin{equation}
		\label{eq:exact_discrete_capacitance}
		C_{p,q}(\etaT)=C_g+C_z|e^{iq}-1|^2
		+C_\phi\left|e^{ip}-1-\etaT(e^{iq}-1)\right|^2 .
	\end{equation}
	In the long-wavelength limit,
	\begin{align}
		\label{eq:long_wave_capacitance}
		C_{p,q}(\etaT)&\simeq C_g+C_zq^2+C_\phi(p-\etaT q)^2
		\nonumber\\
		&=C_g+C_\phi p^2+
		(C_z+C_\phi\etaT^2)q^2-2C_\phi\etaT pq .
	\end{align}
	This is the discrete realization of Eq.~\eqref{eq:q_contraction}; the quantitative correspondence between $(C_\phi,C_z,\etaT)$ and the continuum kernel is established in Sec.~\ref{subsec:dictionary}.
	
	Adding a positive inductive kernel $K_{p,q}$, independent of the twist for simplicity, the normal-mode frequency is
	\begin{equation}
		\label{eq:discrete_frequency}
		\omega_{p,q}(\etaT)=\sqrt{\frac{K_{p,q}}{C_{p,q}(\etaT)}} .
	\end{equation}
	The chiral splitting follows from comparing $p$ and $-p$ at fixed $q$:
	\begin{equation}
		\label{eq:discrete_cap_splitting}
		C_{p,q}(\etaT)-C_{-p,q}(\etaT)
		\simeq -4C_\phi\etaT pq,
	\end{equation}
	and, for weak splitting,
	\begin{equation}
		\label{eq:discrete_frequency_splitting}
		\Delta\omega_{p,q}(\etaT)
		\equiv \omega_{p,q}(\etaT)-\omega_{-p,q}(\etaT)
		\simeq \frac{2C_\phi\etaT pq}{C_{p,q}(0)}\omega_{p,q}(0).
	\end{equation}
	Equation~\eqref{eq:discrete_frequency_splitting} expresses the chiral splitting entirely in terms of measurable circuit parameters.
	
	\subsection{Explicit two-node netlist}
	\label{subsec:netlist}
	
\begin{figure}[t]
		\centering
		\includegraphics[width=0.96\columnwidth]{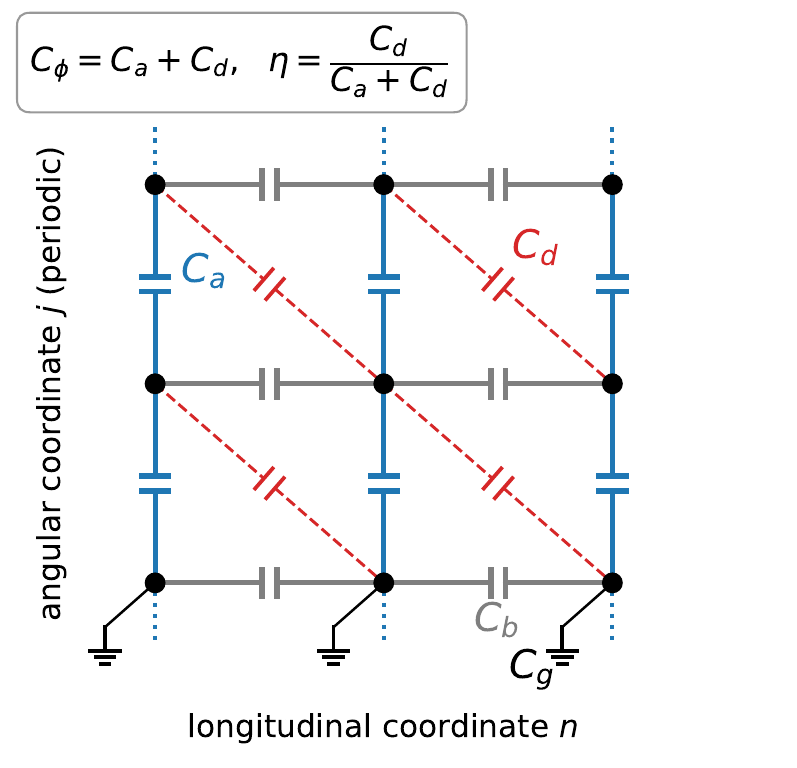}
		\caption{Minimal two-node netlist for the twist capacitance kernel, Eq.~\eqref{eq:netlist_energy}.  Node fluxes $\Phi_{j,n}$ live on an angular synthetic coordinate $j$ (periodic, dotted continuation) and a longitudinal coordinate $n$.  Angular links $C_a$, longitudinal links $C_b$, ground capacitors $C_g$, and diagonal bridges $C_d$ are all ordinary two-terminal capacitors.  The bridges alone generate the chiral cross term $-2C_dpq$ in Eq.~\eqref{eq:netlist_long_wave}, and the twist is the capacitance ratio $\eta=C_d/(C_a+C_d)$.}
		\label{fig:discrete_architecture}
	\end{figure}

	We now show that the spectrum above does not require the three-node coupler of Eq.~\eqref{eq:discrete_TC_square}.  The elementary building blocks are ground capacitors and three families of two-node link capacitors: angular links, longitudinal links, and diagonal bridges,
	\begin{align}
		\label{eq:netlist_energy}
		T_C=\frac{1}{2}\sum_{j,n}\Big[&
		C_g\dot\Phi_{j,n}^2
		+C_a\left(\dot\Phi_{j+1,n}-\dot\Phi_{j,n}\right)^2
		\nonumber\\
		&+C_b\left(\dot\Phi_{j,n+1}-\dot\Phi_{j,n}\right)^2
		\nonumber\\
		&+C_d\left(\dot\Phi_{j+1,n}-\dot\Phi_{j,n+1}\right)^2\Big] ,
	\end{align}
	with $C_g,C_a,C_b,C_d\geq 0$.  Every term is the energy of an ordinary capacitor between two nodes (or node and ground), so the capacitance matrix is automatically real, symmetric, and positive.  In Fourier space,
	\begin{align}
		\label{eq:netlist_symbol}
		C_{p,q}=C_g&+C_a|e^{ip}-1|^2+C_b|e^{iq}-1|^2
		\nonumber\\
		&+C_d|e^{ip}-e^{iq}|^2 ,
	\end{align}
	whose long-wavelength limit is
	\begin{equation}
		\label{eq:netlist_long_wave}
		C_{p,q}\simeq C_g+(C_a+C_d)p^2+(C_b+C_d)q^2-2C_dpq .
	\end{equation}
	The diagonal bridge alone generates the chiral cross term.  Comparing with the expanded form in Eq.~\eqref{eq:long_wave_capacitance} fixes the dictionary
	\begin{equation}
		\label{eq:netlist_dictionary}
		C_\phi=C_a+C_d,
		\qquad
		\etaT=\frac{C_d}{C_a+C_d},
	\end{equation}
	together with $C_z=C_b+C_d-\etaT^2C_\phi$.  The twist is therefore simply the fraction of the total angular capacitance that is routed through the diagonal bridges.  Inverting,
	\begin{equation}
		\label{eq:netlist_inverse}
		C_d=\etaT C_\phi,
		\quad
		C_a=(1-\etaT)C_\phi,
		\quad
		C_b=C_z-\etaT(1-\etaT)C_\phi ,
	\end{equation}
	so any target kernel with $0\leq\etaT\leq1$ is realizable provided
	\begin{equation}
		\label{eq:realizability}
		C_z\geq \etaT(1-\etaT)C_\phi ,
	\end{equation}
	which is guaranteed for all $\etaT\in[0,1]$ whenever $C_z\geq C_\phi/4$.  Negative twists are obtained by replacing the diagonal $(j+1,n)\!\leftrightarrow\!(j,n+1)$ with the antidiagonal $(j,n)\!\leftrightarrow\!(j+1,n+1)$, which flips the sign of the cross term in Eq.~\eqref{eq:netlist_long_wave}.  The exact lattice symbols of Eqs.~\eqref{eq:exact_discrete_capacitance} and \eqref{eq:netlist_symbol} differ at short wavelengths, but they coincide through quadratic order in $(p,q)$, which is the regime where the geometric interpretation applies; in particular the chiral term is identical.  Figure~\ref{fig:discrete_architecture} sketches the minimal architecture.  The parameter $\etaT$ may be swept in situ by implementing the diagonal bridges as tunable capacitive couplers.  Thus the metric twist becomes a circuit-design parameter rather than a literal spatial deformation.
	
	\subsection{Continuum-to-circuit dictionary}
	\label{subsec:dictionary}
	
	The correspondence between the lattice parameters and the continuum kernel of Eq.~\eqref{eq:modal_kernel} can be made quantitative.  For a fixed radial profile $f(r)$, define the radial moments
	\begin{align}
		\label{eq:radial_moments}
		I_{-1}&=\int_0^R \frac{dr}{r}\,|f|^2,
		&
		I_{1}&=\int_0^R dr\,r|f|^2,
		\nonumber\\
		I_{3}&=\int_0^R dr\,r^3|f|^2 .
	\end{align}
	Discarding the twist-independent radial-gradient term, the modal kernel of Eq.~\eqref{eq:modal_kernel} reads
	\begin{align}
		\label{eq:kernel_moments}
		\mathcal{K}_{mk}(\Om)
		&=m^2I_{-1}-2\Om mk\,I_1+\left(I_1+\Om^2I_3\right)k^2
		\nonumber\\
		&=I_{-1}\left(m-\Om\langle r^2\rangle k\right)^2
		+\left[I_1+\Om^2\!\left(I_3-\tfrac{I_1^2}{I_{-1}}\right)\right]k^2 ,
	\end{align}
	where $\langle r^2\rangle=I_1/I_{-1}$ is the mean-square radius with weight $|f|^2/r$.  The Cauchy--Schwarz inequality $I_1^2\leq I_{-1}I_3$ guarantees that both coefficients are positive, so the continuum kernel itself already has the structure of Eq.~\eqref{eq:long_wave_capacitance}.  Identifying $p=2\pi m/N_\phi$ and $q=kd$, with $d$ the longitudinal cell size, and matching Eq.~\eqref{eq:kernel_moments} to Eq.~\eqref{eq:long_wave_capacitance} up to a common positive prefactor set by $\epsilon$ and the mode normalization, gives
	\begin{equation}
		\label{eq:continuum_dictionary}
		C_\phi\propto\left(\frac{N_\phi}{2\pi}\right)^2 I_{-1},
		\qquad
		\etaT=\frac{2\pi}{N_\phi d}\,\Om\langle r^2\rangle ,
	\end{equation}
	and $C_z\propto[I_1+\Om^2(I_3-I_1^2/I_{-1})]/d^2$.  For a mode concentrated in a thin shell at $r=R$, one has $\langle r^2\rangle=R^2$, and for isotropic cells, in which the angular arc $2\pi R/N_\phi$ equals $d$, the dictionary collapses to $\etaT=\Om R$, in exact agreement with the definition of Eq.~\eqref{eq:eta_def}.  The lattice construction therefore reproduces the continuum twist parameter without an adjustable conversion factor.
	
	\section{Chiral spectra from the discrete model}
	\label{sec:chiral_modes}
	
	\subsection{Representative spectra}
	\label{subsec:representative_spectra}
	
	The discrete circuit model turns the formal angular--axial mixing into a directly computable spectrum.  To display the effect, take the dimensionless parameters
	\begin{equation}
		\label{eq:numerical_parameters}
		C_g=1,
		\qquad C_\phi=0.14,
		\qquad C_z=0.06,
	\end{equation}
	and a smooth positive inductive kernel
	\begin{equation}
		\label{eq:inductive_kernel}
		K_{p,q}=K_0+K_\phi\sin^2\left(\frac{p}{2}\right)
		+K_z\sin^2\left(\frac{q}{2}\right),
	\end{equation}
	with $K_0=1$, $K_\phi=0.25$, and $K_z=0.15$.  These numbers are not intended as device optimization; they define a stable representative circuit graph.  The qualitative result does not depend on their precise values.  Note that this parameter set satisfies the netlist realizability condition, Eq.~\eqref{eq:realizability}, for the entire range $\etaT\in[0,1]$: in the dimensionless units of Eq.~\eqref{eq:numerical_parameters}, $\max_{\etaT\in[0,1]}\etaT(1-\etaT)C_\phi=0.035$, while $C_z=0.06$.
	
\begin{figure}[t]
		\centering
		\includegraphics[width=0.96\columnwidth]{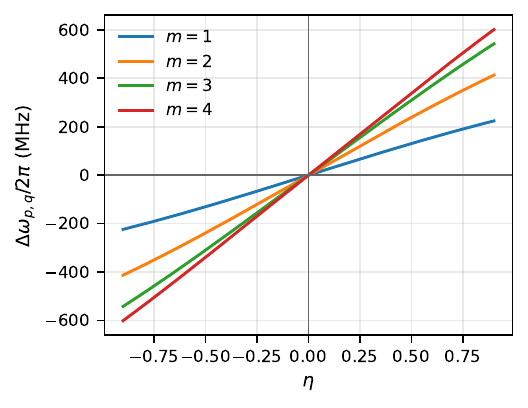}
		\caption{Chiral mode splitting computed from the netlist capacitance symbol, Eq.~\eqref{eq:netlist_symbol}, and the frequency formula, Eq.~\eqref{eq:discrete_frequency}, expressed in SI units ($C_g=50\,\mathrm{fF}$, $L_0=20\,\mathrm{nH}$, base frequency $5.0\,\mathrm{GHz}$).  The plot shows $\Delta\omega_{p,q}=\omega_{p,q}-\omega_{-p,q}$ for fixed $q=0.7$ and representative angular momenta $p=2\pi m/N_\phi$ with $N_\phi=20$.  The response is odd in $\eta$ and increases with the angular sector.}
		\label{fig:discrete_spectrum}
	\end{figure}

	Figure~\ref{fig:discrete_spectrum} shows the resulting chiral frequency splitting for fixed longitudinal momentum $q=0.7$ and several angular sectors, computed from the two-node netlist symbol, Eq.~\eqref{eq:netlist_symbol}, with the components fixed by the dictionary, Eq.~\eqref{eq:netlist_inverse}, and expressed in SI units through the anchoring of Sec.~\ref{subsec:si_estimates}.  The splitting is odd in the twist, reverses under $p\to -p$, and grows with the angular index.  This is the mesoscopic circuit signature of the helicoidal geometry.
	
\begin{figure*}[t]
		\centering
		\includegraphics[width=0.98\textwidth]{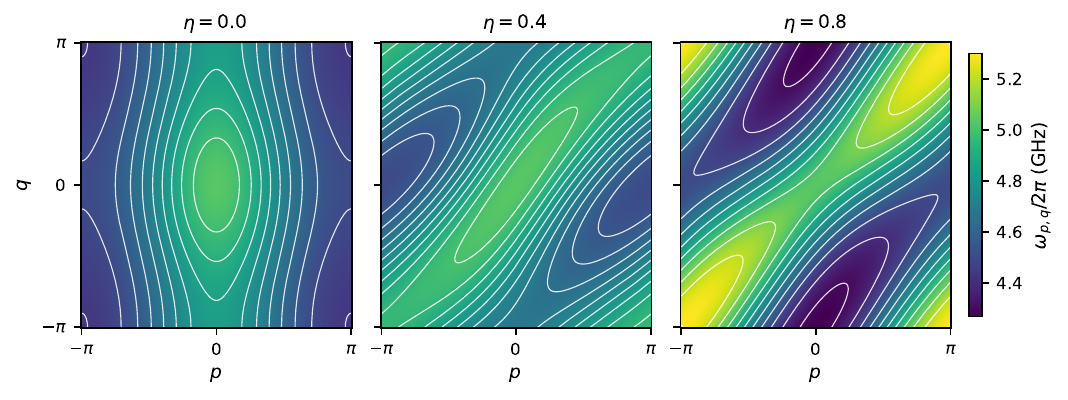}
		\caption{Iso-frequency contours of the netlist dispersion $\omega_{p,q}$ [Eqs.~\eqref{eq:netlist_symbol} and \eqref{eq:discrete_frequency}] across the synthetic Brillouin zone, in SI units, for $\eta=0$, $0.4$, and $0.8$.  The twist shears the contours along $p=\eta q$, breaking the $p\to-p$ symmetry at fixed $q$ while preserving the helical symmetry $(p,q)\to(-p,-q)$.}
		\label{fig:isofreq}
	\end{figure*}

	The same shear is visible at the level of the full dispersion.  Figure~\ref{fig:isofreq} shows iso-frequency contours of $\omega_{p,q}$ across the synthetic Brillouin zone for increasing twist.  At $\etaT=0$ the contours are symmetric under $p\to-p$ at fixed $q$; the twist tilts them along the direction $p=\etaT q$, in close analogy with the shear of constant-energy surfaces in strained or twisted crystals.  The doublet splittings of Fig.~\ref{fig:discrete_spectrum} are one-dimensional cuts of this global deformation.
	
	\subsection{Symmetry, and relation to synthetic gauge fields}
	\label{subsec:gauge_relation}
	
	The important point is not the particular numerical scale, but the symmetry.  A purely longitudinal capacitance gives a quadratic response in $\etaT$.  A twist graph gives a linear, chiral response because the capacitance eigenvalue depends on $p-\etaT q$.  In a circuit-QED experiment, this splitting would appear as a twist-dependent detuning between counter-propagating synthetic modes, and it could be read out spectroscopically.
	
	It is worth distinguishing this mechanism from a synthetic magnetic flux.  The shifted symbol $C_{p,q}\propto(p-\etaT q)^2$ resembles minimal coupling to a gauge potential, but it acts in the \emph{reactive} (capacitive) sector: all couplings in Eq.~\eqref{eq:netlist_energy} are real and positive, no hopping phases appear, and time-reversal symmetry is preserved.  Indeed, the symbols in Eqs.~\eqref{eq:exact_discrete_capacitance} and \eqref{eq:netlist_symbol} obey $C_{p,q}=C_{-p,-q}$, hence $\omega_{p,q}=\omega_{-p,-q}$, so the splitting is helical rather than magnetic: it is odd under $p\to-p$ at fixed $q$, vanishes at $q=0$, and is even under the simultaneous reversal $(p,q)\to(-p,-q)$.  This contrasts with time-reversal-broken photon lattices, where chirality originates from complex hopping amplitudes generated by parametric modulation or ferrite elements and $\omega_{p,q}\neq\omega_{-p,-q}$ in general \cite{Koch2010,Roushan2017}.  The twist circuit is therefore closer in spirit to metric engineering \cite{Kollar2019,Boettcher2020} than to gauge-flux engineering: it realizes a screw-like deformation of the synthetic space itself.
	
	\subsection{Exact diagonalization of a finite graph, and the role of longitudinal closure}
	\label{subsec:finite_graph}
	
\begin{figure}[t]
		\centering
		\includegraphics[width=0.96\columnwidth]{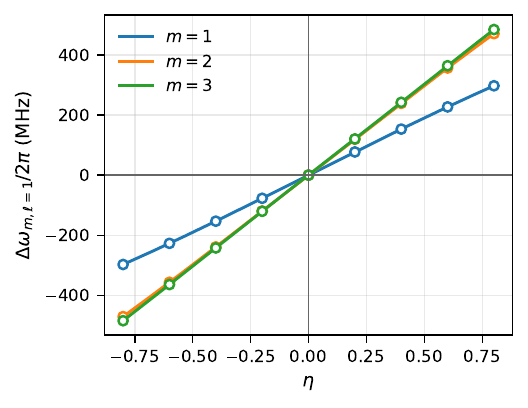}
		\caption{Chiral splitting from exact real-space diagonalization of a finite $12\times10$ netlist graph [Eq.~\eqref{eq:netlist_energy}] with both synthetic directions closed (torus).  Symbols: doublet splittings $(m,\ell=1)$ obtained from the generalized eigenproblem, Eq.~\eqref{eq:generalized_eigenproblem}.  Lines: analytic prediction from Eqs.~\eqref{eq:netlist_symbol} and \eqref{eq:discrete_frequency}.  Frequencies are expressed in SI units using $C_g=50\,\mathrm{fF}$ and $L_0=20\,\mathrm{nH}$.  With open longitudinal boundaries the splitting vanishes identically (see text).}
		\label{fig:finite_graph}
	\end{figure}

	The Fourier analysis above assumes periodic boundary conditions.  In real space, the normal modes of a finite netlist solve the generalized eigenvalue problem
	\begin{equation}
		\label{eq:generalized_eigenproblem}
		\omega^2\,C\,\bm v = L^{-1}\bm v ,
	\end{equation}
	with $C$ and $L^{-1}$ the $N_\phi N_z\times N_\phi N_z$ capacitance and inverse-inductance matrices of the graph, the latter built from ground inductors on every node and link inductors along both directions, reproducing Eq.~\eqref{eq:inductive_kernel} in the periodic case.  We have diagonalized Eq.~\eqref{eq:generalized_eigenproblem} numerically for a $N_\phi\times N_z=12\times10$ netlist with both directions closed (a synthetic torus).  In a circuit this closure does not require bending the chip into a real torus; it is implemented by engineered interconnects that wire column $n=N_z-1$ back to column $n=0$.  In practice, such closure links require layout-compatible crossovers, controlled parasitic capacitances, and calibration of the bridge capacitances, so they should be regarded as a concrete design requirement rather than as a cost-free geometric operation.  The exact eigenfrequencies, extracted sector by sector through projection onto $e^{i(pj+qn)}$, agree with the analytic symbol of Eq.~\eqref{eq:netlist_symbol} up to numerical roundoff, with relative deviations below $10^{-14}$ in the finite graph used below, and the chiral splittings are linear and odd in $\etaT$, as shown in Fig.~\ref{fig:finite_graph}.
	
	The same computation reveals a feature that sharpens the experimental requirements.  If the longitudinal direction is left \emph{open}, the sectors $+m$ and $-m$ are exactly degenerate for every $\etaT$: numerically the splitting vanishes to machine precision, and analytically this follows because, in each angular sector, the diagonal bridge only endows the effective one-dimensional longitudinal chain with a complex hopping phase, which is a removable gauge on an open chain.  This is the circuit analogue of the fact that the continuum coupling $-2\Om mk$ averages to zero for longitudinal standing waves, which contain $+k$ and $-k$ in equal measure.  The chiral splitting therefore requires closed longitudinal circulation.  In practice this means either closing the synthetic torus through calibrated interconnects, or probing an open lattice with traveling waves, where the twist appears as a direction-dependent transmission phase rather than as an eigenfrequency splitting.
	
	\subsection{Simulated chiral spectroscopy}
	\label{subsec:spectroscopy}
	
\begin{figure}[t]
		\centering
		\includegraphics[width=0.96\columnwidth]{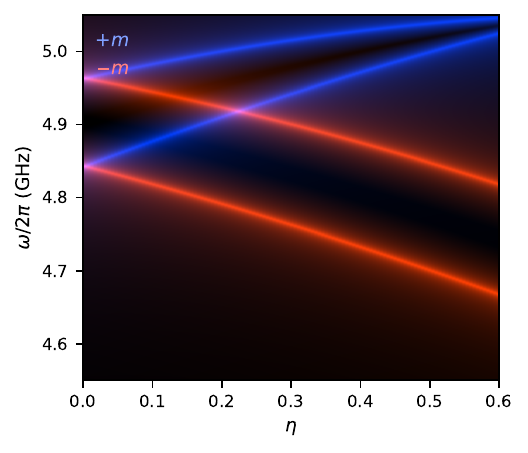}
		\caption{Simulated chiral spectroscopy of the $12\times10$ torus, Eq.~\eqref{eq:response}.  A single node is driven and the response is read out by helical phase-matched combs $\sum_{m=1,2}e^{i(\pm m\,2\pi j/N_\phi+2\pi n/N_z)}$ selecting the co-rotating ($+m$, blue) and counter-rotating ($-m$, red) channels; brightness encodes $\log_{10}|\chi_\pm|$ with uniform linewidth $\kappa/2\pi\simeq4\,\mathrm{MHz}$.  The chiral doublets open linearly in $\etaT$.}
		\label{fig:transmission}
	\end{figure}

	The chiral fan can be observed directly in the driven response.  We model spectroscopy of the $12\times10$ torus by the damped linear response
	\begin{equation}
		\label{eq:response}
		\bm\chi(\omega)=\left(L^{-1}-\omega^2C-i\omega G\right)^{-1}
		\bm{\mathcal{I}},
	\end{equation}
	with a drive current $\bm{\mathcal{I}}$ applied to a single node and a uniform shunt conductance $G$ chosen so that every mode has linewidth $\kappa/2\pi\simeq4\,\mathrm{MHz}$.  Detection uses helical phase-matched combs, $d_\pm\propto\sum_{j,n}e^{i(\pm m\,2\pi j/N_\phi+2\pi n/N_z)}$, which project the response onto a single chirality.  Figure~\ref{fig:transmission} overlays the two channels: the co-rotating branches (blue) and counter-rotating branches (red) separate linearly in $\etaT$, forming the chiral fan that a vector network analyzer would record.
	
	An instructive subtlety fixes the design of the probe.  A detection comb that resolves only the angular phase is blind to the twist: on the torus the frequency sets $\{\omega_{m,\ell}\}_\ell$ and $\{\omega_{-m,\ell}\}_\ell$ coincide, being exchanged by $\ell\to-\ell$, so the two channels see identical resonance positions.  Only probes that are phase matched in \emph{both} synthetic directions---helical probes---separate the branches.  This is the response-function manifestation of the holonomy identified in Sec.~\ref{subsec:finite_graph}: the chirality is a property of circulating helical modes, not of any local or single-direction observable.
	
	\subsection{Robustness to fabrication disorder}
	\label{subsec:disorder}
	
\begin{figure}[t]
		\centering
		\includegraphics[width=0.96\columnwidth]{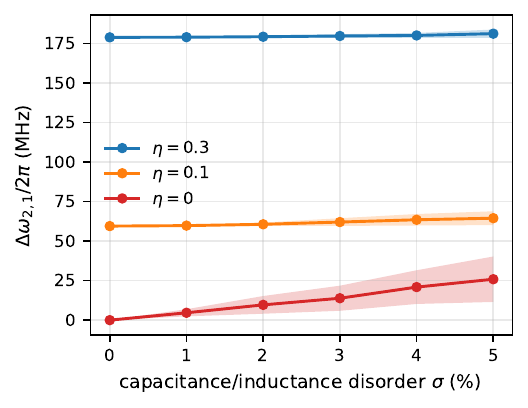}
		\caption{Disorder robustness of the chiral splitting.  Mean $\pm$ one standard deviation of the $(2,1)$ doublet gap over $120$ realizations of independent multiplicative disorder of strength $\sigma$ on every capacitor and inductor of the $12\times10$ torus.  The finite-twist curves remain well separated from the disorder-induced noise floor for the reference parameters; the $\etaT=0$ curve measures the disorder-induced coupling $2J_{\rm dis}$.}
		\label{fig:disorder}
	\end{figure}

	Fabrication tolerance is the natural experimental concern for a lattice of nominally identical capacitors.  We add independent Gaussian multiplicative disorder of fractional strength $\sigma$ to \emph{every} capacitor and inductor of the $12\times10$ torus and evaluate the $(m,\ell)=(2,1)$ doublet gap by projecting the generalized eigenproblem onto the doublet subspace ($120$ realizations per point).  Figure~\ref{fig:disorder} shows the result.  The chiral splitting is statistically robust in this disorder model: at $\etaT=0.3$ it moves from $179$ to $181\pm3\,\mathrm{MHz}$ under $\sigma=5\%$ disorder.  This should not be interpreted as an exact first-order protection against arbitrary reactive disorder.  The partners compared in the chiral splitting are $(p,q)$ and $(-p,q)$, whereas the complex-conjugate partner of $(p,q)$ is $(-p,-q)$.  Rather, the numerical result reflects the fact that independent real capacitance and inductance fluctuations mainly produce common-mode shifts and a comparatively small disorder-induced coupling within the chosen doublet.  The latter is directly visible as the level-repulsion gap that opens at $\etaT=0$, $2J_{\rm dis}/2\pi\simeq5\,\mathrm{MHz}$ at $\sigma=1\%$ and $26\,\mathrm{MHz}$ at $\sigma=5\%$.  The deterministic twist splitting remains faithful as long as $\Delta\omega\gg2J_{\rm dis}$, a condition met with margin for the reference parameters and percent-level tolerances.
	
	\subsection{Measurability in SI units}
	\label{subsec:si_estimates}
	
	To translate the dimensionless spectra into laboratory numbers, take $C_g=50\,\mathrm{fF}$, so that $C_\phi=7\,\mathrm{fF}$ and $C_z=3\,\mathrm{fF}$, all standard circuit-QED component scales; the diagonal bridges then span $C_d=\etaT C_\phi\leq 7\,\mathrm{fF}$.  Ground inductors $L_0\simeq 20\,\mathrm{nH}$ set the base frequency $\omega/2\pi=1/(2\pi\sqrt{L_0C_g})\simeq 5.0\,\mathrm{GHz}$, in the standard circuit-QED band.  With this anchoring, the netlist symbol gives, for the $m=2$ doublet at $q=0.7$ ($N_\phi=20$), a splitting $\Delta\omega/2\pi\simeq145\,\mathrm{MHz}$ at $\etaT=0.3$ and $\simeq24\,\mathrm{MHz}$ even at $\etaT=0.05$; the corresponding torus doublet $(m=2,\ell=1)$ of the $12\times10$ graph gives $179$ and $30\,\mathrm{MHz}$, respectively.  Both scales exceed typical linewidths of superconducting microwave resonators, $\kappa/2\pi\sim0.1$--$1\,\mathrm{MHz}$ \cite{Blais2021}, by one to three orders of magnitude.  The chiral splitting should therefore be resolvable by standard transmission spectroscopy, provided that diagonal bridge capacitances, closure links, and parasitic capacitances are calibrated well enough that disorder-induced mode mixing remains below the deterministic splitting.  Table~\ref{tab:parameters} collects the device parameters and the predicted observables in one place.
	
	\begin{table}[t]
		\caption{\label{tab:parameters}Summary of reference device parameters and predicted observables for the $12\times10$ toroidal netlist.  Literature references are cited directly in the rows associated with representative circuit-QED scales; entries marked as ``this work'' are predictions of the present model.  Splittings and Kerr coefficients refer to the $(m,\ell)=(2,1)$ chiral doublet.}
		\begin{ruledtabular}
			\begin{tabular}{lll}
				Quantity & Symbol & Value/source \\
				\colrule
				Ground capacitance & $C_g$ & $50\,\mathrm{fF}$~\cite{Vool2017,Blais2021} \\
				Angular capacitance & $C_\phi$ & $7\,\mathrm{fF}$~\cite{Vool2017,Blais2021} \\
				Longitudinal capacitance & $C_z$ & $3\,\mathrm{fF}$~\cite{Vool2017,Blais2021} \\
				Diagonal bridge & $C_d=\etaT C_\phi$ & $0$--$7\,\mathrm{fF}$~\cite{Vool2017,Blais2021} \\
				Ground inductance & $L_0$ & $20\,\mathrm{nH}$~\cite{Blais2021} \\
				Base frequency & $\omega/2\pi$ & $5.0\,\mathrm{GHz}$~\cite{Wallraff2004,Blais2021} \\
				Assumed linewidth & $\kappa/2\pi$ & $0.1$--$1\,\mathrm{MHz}$~\cite{Blais2021} \\
				\colrule
				Chiral splitting ($\etaT=0.1$) & $\Delta\omega_{2,1}/2\pi$ & $60\,\mathrm{MHz}$; this work \\
				Chiral splitting ($\etaT=0.3$) & $\Delta\omega_{2,1}/2\pi$ & $179\,\mathrm{MHz}$; this work \\
				Disorder floor ($\sigma=1\%$) & $2J_{\rm dis}/2\pi$ & $5\,\mathrm{MHz}$; this work \\
				Junction energy & $E_J/h$ & $8.2\,\mathrm{GHz}$~\cite{Koch2007} \\
				Charging energy & $\EC/h$ & $0.39\,\mathrm{GHz}$~\cite{Koch2007} \\
				Self-Kerr & $\bar K^{\rm Kerr}/2\pi$ & $-2.6\,\mathrm{MHz}$~\cite{Nigg2012,Kirchmair2013} \\
				Cross-Kerr & $\chi_{+-}/2\pi$ & $-5.2\,\mathrm{MHz}$~\cite{Nigg2012,Kirchmair2013} \\
				Kerr asymmetry ($\etaT=0.3$) & $\delta K/2\pi$ & $-0.19\,\mathrm{MHz}$; this work \\
				EP coupling & $J/2\pi$ & $10\,\mathrm{MHz}$~\cite{Song2024} \\
			\end{tabular}
		\end{ruledtabular}
	\end{table}
	
	\section{Twist domain walls and heterojunctions}
	\label{sec:domain_walls}
	
	Because the twist is a per-link design parameter, it can be textured: $\etaT$ may take different values in different regions of the same lattice.  We consider longitudinal profiles $\etaT(n)$ on a $12\times40$ torus and diagonalize the full real-space problem.  Two textures are particularly instructive.
	
	The first is a symmetric domain wall, $\etaT=+\etaT_0$ on one half of the torus and $-\etaT_0$ on the other, with $\etaT_0=0.8$.  In the ideal symmetric model the twist fluxes of the two domains cancel: in each angular sector the accumulated bridge phase winds by equal and opposite amounts, so the total holonomy vanishes and the net chiral response is expected to be strongly suppressed despite the lattice being maximally twisted almost everywhere.  The simulated chiral response confirms this cancellation mechanism: the integrated asymmetry between the co- and counter-rotating detection channels, $\mathcal{A}=\int d\omega\,\big||\chi_+|-|\chi_-|\big|\,/\int d\omega\,(|\chi_+|+|\chi_-|)$, drops from $0.11$ for the uniform lattice to $0.014$ for the wall, with the resonance positions of the two channels coinciding within the numerical resolution.  The small residual asymmetry reflects amplitude differences associated with the finite probe and finite-size implementation, not a robust frequency splitting.  A twist texture with zero net winding is therefore nearly achiral for these ideal chiral probes, a testable statement of the holonomy-like character of the twist rather than a claim of universal invisibility.
	
\begin{figure*}[t]
		\centering
		\includegraphics[width=0.98\textwidth]{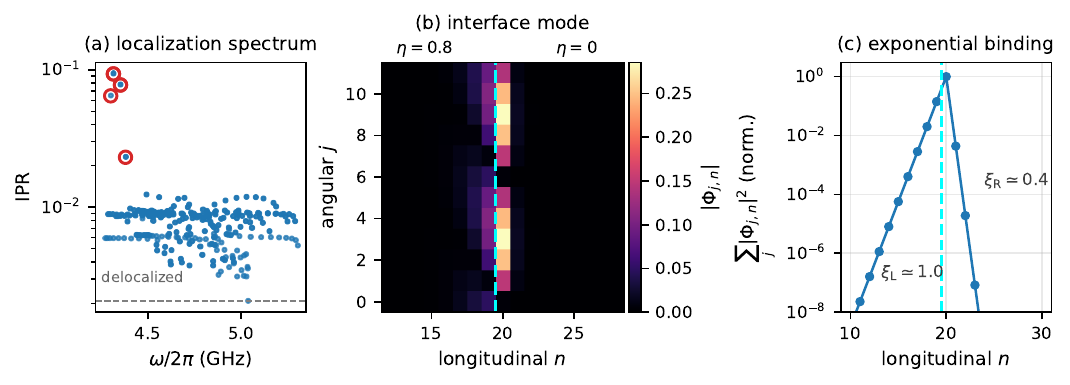}
		\caption{Interface modes of a twist heterojunction ($\etaT=0.8$ for $n<20$, $\etaT=0$ for $n\geq20$, on a $12\times40$ torus).  (a) Inverse participation ratio of all $480$ modes versus frequency; red circles mark the interface-bound modes split off below the bulk band, and the dashed line marks the uniform-lattice value $1/N$.  (b) Amplitude map $|\Phi_{j,n}|$ of the most localized mode near the wall (dashed cyan line): a ring of excitation bound to the bridge-terminated interface, with the twisted and untwisted domains indicated.  (c) Longitudinal profile of the same mode on a logarithmic scale, showing exponential binding with decay lengths $\xi_{\rm L}\simeq1.0$ and $\xi_{\rm R}\simeq0.4$ lattice cells.}
		\label{fig:domain_wall}
	\end{figure*}

	The second texture is a twist heterojunction, $\etaT=\etaT_0$ on one half and $\etaT=0$ on the other.  Here the net winding is nonzero and, in addition, the junction breaks the longitudinal homogeneity of the netlist components.  Exact diagonalization reveals a family of modes bound to the interface, split off below the bulk band: their inverse participation ratio, ${\rm IPR}=\sum_i|\Phi_i|^4/(\sum_i|\Phi_i|^2)^2$, exceeds the uniform-lattice value $1/N$ by a factor of $\sim45$, and their amplitude decays exponentially away from the bridge-terminated wall, with decay lengths $\xi_{\rm L}\simeq1.0$ lattice cells on the twisted side and $\xi_{\rm R}\simeq0.4$ on the untwisted side, forming a ring of localized excitations at $4.30$--$4.38\,\mathrm{GHz}$, roughly $100\,\mathrm{MHz}$ below the band edge and hence individually addressable in spectroscopy (Fig.~\ref{fig:domain_wall}).  The two walls of the profile are inequivalent---one terminates the diagonal bridges, the other does not---and the bound modes attach to the bridge-terminated wall.  These states should be understood as interface and termination modes of the engineered netlist, not as universally protected topological modes.  Twist heterojunctions nevertheless provide localized, spectrally isolated probes of the twist texture, in the spirit of interface states at heterojunctions in semiconductor physics.
	
	\section{Twist-controlled Kerr nonlinearities}
	\label{sec:kerr}
	
	The results so far concern the linear spectrum.  The twist also imprints itself on the \emph{interactions} once Josephson nonlinearity is introduced.  The minimal construction replaces every ground inductor by a Josephson junction with the same linear inductance, $L_J=L_0$, i.e., $E_J=\varphi_0^2/L_0$ with $\varphi_0=\hbar/2e$: the linear spectrum of Secs.~\ref{sec:discrete_circuit}--\ref{sec:chiral_modes} is unchanged, while the quartic term of each junction generates Kerr interactions.  Expanding $-E_J\cos\hat\varphi_i$ to fourth order and normal ordering in the normal-mode basis $\hat\Phi_i=\sum_k x_{ik}\sqrt{\hbar/2\omega_k}\,(\hat a_k+\hat a_k^\dagger)$, with $x^TCx=\openone$, gives the standard black-box form \cite{Nigg2012}
	\begin{equation}
		\label{eq:kerr_hamiltonian}
		\frac{\hat H_{\rm nl}}{\hbar}=\sum_k\frac{K_k}{2}\,
		\hat a_k^{\dagger2}\hat a_k^2
		+\sum_{k<l}\chi_{kl}\,\hat n_k\hat n_l ,
	\end{equation}
	with
	\begin{equation}
		\label{eq:kerr_participation}
		K_k=-\frac{E_J}{2\hbar}\sum_i\varphi_{ik}^4,
		\qquad
		\chi_{kl}=-\frac{E_J}{\hbar}\sum_i\varphi_{ik}^2\varphi_{il}^2 ,
	\end{equation}
	where $\varphi_{ik}=x_{ik}\sqrt{\hbar/2\omega_k}/\varphi_0$ is the zero-point phase of mode $k$ across junction $i$.  Bloch modes have uniform participation, $|x_{ik}|^2=1/(NC_{p,q})$, so the sums collapse to closed forms:
	\begin{align}
		\label{eq:kerr_closed}
		K_{p,q}^{\rm Kerr}
		&=-\frac{\EC}{\hbar}\,
		\frac{C_gK_0}{N\,C_{p,q}K_{p,q}},
		\nonumber\\
		\chi_{kl}
		&=-2\sqrt{\left|K_k^{\rm Kerr}\right|
			\left|K_l^{\rm Kerr}\right|}\, .
	\end{align}
	with $\EC=e^2/2C_g$.  For a single junction ($N=1$, $C_{p,q}\to C_\Sigma$, $K_{p,q}\to K_0$) this reproduces the transmon anharmonicity $K=-\EC/\hbar$, as it must.
	
\begin{figure}[t]
		\centering
		\includegraphics[width=0.96\columnwidth]{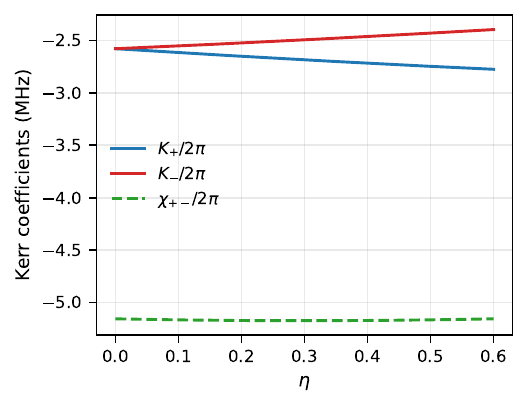}
		\caption{Twist-controlled Kerr coefficients of the $(2,1)$ chiral doublet for the parameters of Sec.~\ref{subsec:si_estimates}, with every ground inductor replaced by a Josephson junction ($E_J/\EC\simeq21$).  The self-Kerr coefficients $K_\pm$ of the counter-rotating partners split linearly in $\etaT$ [Eq.~\eqref{eq:kerr_asymmetry}], while the cross-Kerr $\chi_{+-}$ stays an order of magnitude above typical linewidths, enabling photon-number-resolved, chirality-selective dispersive shifts.}
		\label{fig:kerr}
	\end{figure}

	The chiral content of Eq.~\eqref{eq:kerr_closed} follows from the fact that $K_{p,q}$ is even in $p$ while $C_{p,q}$ is not: the doublet partners acquire \emph{different} self-Kerr coefficients,
	\begin{equation}
		\label{eq:kerr_asymmetry}
		\delta K\equiv K_+^{\rm Kerr}-K_-^{\rm Kerr}
		\simeq\bar K^{\rm Kerr}\,\frac{4C_\phi\etaT pq}{C_{p,q}} ,
	\end{equation}
	a twist-controlled anharmonicity asymmetry, linear and odd in $\etaT$.  For the device parameters of Sec.~\ref{subsec:si_estimates} one has $\EC/h=387\,\mathrm{MHz}$ and $E_J/h=8.2\,\mathrm{GHz}$, so each junction is individually in a moderately transmon-like regime ($E_J/\EC\simeq21$), and the $(2,1)$ doublet of the $12\times10$ torus acquires $\bar K^{\rm Kerr}/2\pi\simeq-2.6\,\mathrm{MHz}$, $\chi_{+-}/2\pi\simeq-5.2\,\mathrm{MHz}$, and $\delta K/2\pi\simeq-190\,\mathrm{kHz}$ at $\etaT=0.3$ (Fig.~\ref{fig:kerr}).  The cross-Kerr exceeds typical linewidths by an order of magnitude, placing the doublet in the photon-number-resolved regime \cite{Kirchmair2013}: a single photon in the counter-rotating mode shifts the co-rotating resonance by several linewidths, enabling chirality-resolved dispersive readout, while $\delta K$ is directly measurable by comparing the two-photon shifts of the two branches.  The twist thus controls not only the linear spectrum but also the interaction sector of the quantum theory.
	
	Two validity remarks are in order.  First, the perturbative treatment requires $|K^{\rm Kerr}|,|\chi|\ll\Delta\omega$ and small zero-point phases; these conditions are satisfied for the reference parameters, but a device-specific design should verify them after electromagnetic extraction of the mode capacitances and junction participations.  Second, exactly degenerate partners $\{(m,\ell),(-m,-\ell)\}$ also acquire resonant pair-exchange quartic terms of the same order; these act within each degenerate pair and do not affect the $\pm m$ doublet at fixed $\ell$, which is split by $\Delta\omega\gg|\chi|$ and therefore purely dispersive.
	
	\section{Non-Hermitian two-mode sector}
	\label{sec:nonhermitian}
	
	Open superconducting circuits provide a natural setting for non-Hermitian dynamics and exceptional-point physics \cite{Miri2019,Ashida2020,Bergholtz2021}.  Exceptional points have been observed both in single dissipative superconducting qubits \cite{Naghiloo2019} and in on-chip superconducting quantum circuits in the few-photon regime \cite{Song2024}.  The discrete twist circuit supplies a concrete two-mode subspace for such a construction: the pair $(p,q)$ and $(-p,q)$.
	
\begin{figure}[t]
		\centering
		\includegraphics[width=0.96\columnwidth]{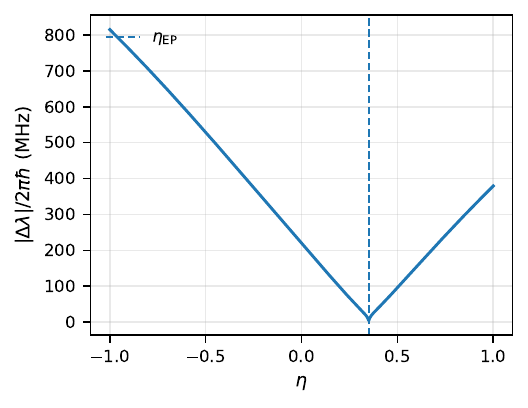}
		\caption{Representative exceptional-point control in the chiral two-mode model, in SI units ($J/2\pi\simeq10\,\mathrm{MHz}$).  The plotted gap follows Eq.~\eqref{eq:ep_condition} with $|\delta\kappa|=4J$ and a twist-tunable real detuning obtained from the discrete chiral spectrum.  The gap closes when the twist compensates the static offset $\delta\omega_0$.}
		\label{fig:ep_control}
	\end{figure}

	The two ingredients beyond the closed lattice are a coherent coupling $J$ and asymmetric decay rates $\kappa_\pm$, and both admit concrete circuit implementations.  A static azimuthal modulation of the ground capacitance, $\delta C_g\cos(2mj\,2\pi/N_\phi)$, Bragg-couples the sectors $+m$ and $-m$ directly and produces a real reciprocal $J$ proportional to $\delta C_g$; weak fabrication disorder produces the same coupling uncontrolled, so a deliberate modulation dominating the disorder scale is preferable.  Asymmetric decay is obtained by coupling one row of nodes to an output transmission line through a chain of coupling capacitors with phases matched to the running wave $e^{ipj}$: such a phase-matched (traveling-wave) coupler overlaps constructively with the co-rotating mode and destructively with the counter-rotating one, yielding $\kappa_+\neq\kappa_-$, in direct analogy with directional couplers in waveguide QED.  Projecting onto the chiral doublet then gives
	\begin{equation}
		\label{eq:heff}
		H_{\rm eff}(\etaT)=\hbar
		\begin{pmatrix}
			\omega_{p,q}(\etaT)-i\kappa_+/2 & J \\
			J & \omega_{-p,q}(\etaT)-i\kappa_-/2
		\end{pmatrix} .
	\end{equation}
	The eigenvalue coalescence condition is
	\begin{equation}
		\label{eq:ep_condition}
		\left[\delta\omega(\etaT)-\frac{i}{2}\delta\kappa\right]^2+4J^2=0,
	\end{equation}
	where
	\begin{equation}
		\label{eq:detunings}
		\delta\omega(\etaT)=\omega_{p,q}(\etaT)-\omega_{-p,q}(\etaT)+\delta\omega_0,
		\qquad
		\delta\kappa=\kappa_+-\kappa_- .
	\end{equation}
	For real reciprocal coupling $J$, an exceptional point occurs when $\delta\omega(\etaT)=0$ and $|\delta\kappa|=4J$.  Since the first term in $\delta\omega(\etaT)$ is twist tunable, the EP position is approximately
	\begin{equation}
		\label{eq:eta_ep}
		\etaT_{\rm EP}\simeq -\frac{\delta\omega_0}{v_{p,q}},
		\qquad
		v_{p,q}=\left.\frac{\partial}{\partial\etaT}
		\left[\omega_{p,q}(\etaT)-\omega_{-p,q}(\etaT)\right]\right|_{\etaT=0} .
	\end{equation}
	Thus the non-Hermitian control is not appended phenomenologically: it uses the chiral splitting produced by the same discrete capacitance matrix.  Figure~\ref{fig:ep_control} shows the eigenvalue gap along a one-dimensional twist sweep at fixed $|\delta\kappa|=4J$, closing at $\etaT_{\rm EP}$.  The two-mode reduction is valid as long as $|\delta\omega|$, $\kappa_\pm$, and $J$ remain small compared with the spacing to the neighboring lattice modes, a condition easily met by the parameters of Sec.~\ref{subsec:si_estimates}, where doublet splittings of tens of MHz coexist with intermode spacings of hundreds of MHz.
	
\begin{figure*}[t]
		\centering
		\includegraphics[width=0.9\textwidth]{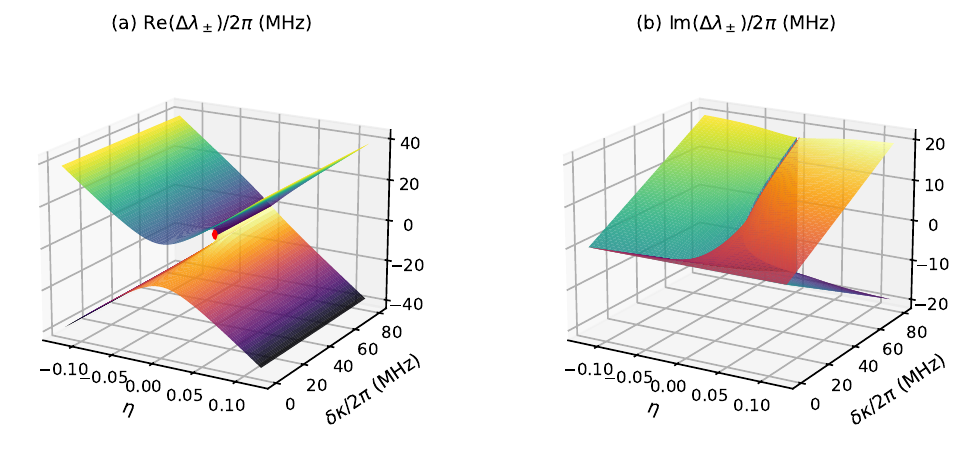}
		\caption{Riemann-sheet structure of the chiral two-mode sector over the $(\etaT,\delta\kappa)$ control plane, for $J/2\pi=10\,\mathrm{MHz}$ and the twist-to-detuning slope of the $(m,q)=(3,0.7)$ doublet.  (a) Real and (b) imaginary parts of the eigenvalue difference $\Delta\lambda_\pm$ of Eq.~\eqref{eq:heff}.  The exceptional point (red dot) sits at $\delta\omega(\etaT)=0$, $\delta\kappa=4J$, where the two sheets coalesce.}
		\label{fig:ep_riemann}
	\end{figure*}

	Figure~\ref{fig:ep_riemann} displays the global structure of the complex eigenvalue difference over the two-dimensional control plane spanned by the twist and the dissipation asymmetry.  The real parts form the characteristic double cone that closes along the compensated-detuning line, the imaginary parts bifurcate beyond it, and the two behaviors exchange at the exceptional point $\delta\kappa=4J$.  Because one control axis is the twist itself, encircling the exceptional point---with the associated eigenvalue braiding---amounts to sweeping a capacitive coupler against a dissipation asymmetry, both standard operations in superconducting circuits \cite{Naghiloo2019,Song2024}.
	
	\section{Discussion and outlook}
	\label{sec:discussion}
	
	The discrete construction is the crucial bridge between geometry and implementation.  It shows that the helicoidal kernel is not only a continuum analogy: a finite graph built exclusively from positive two-node capacitors reproduces the long-wavelength structure $C_{p,q}\sim C_g+C_zq^2+C_\phi(p-\eta q)^2$, with the twist given by the fraction of angular capacitance routed through diagonal bridges, Eq.~\eqref{eq:netlist_dictionary}, subject only to the mild realizability condition $C_z\geq C_\phi/4$.  The twist is then encoded by circuit layout or tunable bridge capacitances, in the same spirit that curved lattices in circuit QED encode geometry through connectivity rather than real spatial curvature \cite{Kollar2019,Boettcher2020}.  A device implementation would still require a full electromagnetic extraction of parasitic capacitances, crossover-induced shifts, and closure-link imperfections; these effects calibrate the realized parameters rather than invalidate the positivity or quantizability of the proposed matrix.  The present work is not intended as a complete microwave-layout proposal; rather, it identifies the positive capacitance matrix, spectral signatures, and parameter ranges that a subsequent electromagnetic design should target.
	
	The classical calculation remains useful because it identifies which geometric responses are even or odd in the twist.  The longitudinal coordinate capacitor gives a quadratic shift of $C_z$, which translates into a quadratic shift of $E_C$, $\omega_{01}$, and the transmon anharmonicity.  The discrete angular--axial sector gives the stronger result: a linear splitting between counter-rotating synthetic modes, confirmed here by exact diagonalization of a finite toroidal graph and estimated at tens to hundreds of MHz for standard device parameters.  The requirement of closed longitudinal circulation, identified in Sec.~\ref{subsec:finite_graph}, is itself a physically meaningful prediction: the twist is a holonomy-like property of circulating synthetic modes, not a local level shift.  The construction is therefore a progression from geometric capacitance to Hamiltonian engineering and finally to chiral circuit spectra, with each step expressed in fabricable circuit elements.
	
	Several extensions are immediate.  First, one can move beyond the lumped-element idealization and extract the black-box capacitance matrix of a concrete layout, including parasitics, from electromagnetic simulation.  Second, one can combine the chiral doublet with the engineered loss channels of Sec.~\ref{sec:nonhermitian} to encircle the exceptional point dynamically and map out its topology, including the associated chiral state transfer.  Third, a time-dependent twist $\etaT(t)$, modulated at the doublet splitting through tunable bridge couplers, would convert the counter-rotating partners coherently, opening a route to twist-driven parametric processes.  These steps would move the proposal from a minimal analytic model to a device-specific implementation.
	
	\section{Conclusions}
	\label{sec:conclusions}
	
	We have developed a route for implementing a torsionless helicoidal synthetic geometry in mesoscopic quantum circuits.  The metric determines a capacitance kernel whose angular--axial sector carries the chiral term $-2\Om mk$, and this kernel is realized by a finite graph built exclusively from positive two-node capacitors, with the twist set by the ratio of diagonal to total angular capacitance and connected to the continuum without adjustable conversion factors.  Upon quantization, the capacitance matrix enters the Hamiltonian through $C^{-1}(\eta)$ and splits counter-rotating synthetic modes by tens to hundreds of MHz for standard device parameters, provided the synthetic longitudinal direction is closed into a torus; for open boundaries the splitting vanishes by a gauge argument, making closed circulation an essential design requirement.  The splitting is resolvable by helical phase-matched spectroscopy and remains above the disorder-induced coupling floor in the calibrated percent-level disorder model considered here.  Twist textures extend the platform beyond uniform geometries: symmetric domain walls strongly suppress the net chiral response in the ideal model, while heterojunctions bind spectrally isolated interface modes.  Josephson nonlinearity promotes the twist to a control knob of the interaction sector, splitting the doublet self-Kerr coefficients by hundreds of kHz while the cross-Kerr remains photon-number resolving.  The same chiral doublet supports a non-Hermitian extension in which the twist acts as a detuning knob for exceptional-point control.

	\section*{Data availability}
	
	The analytic formulas needed to reproduce the figures are given in the text.  The numerical data are generated from the capacitance symbol, Eq.~\eqref{eq:netlist_symbol}, and from the real-space stamping rules described in the finite-graph construction.  The numerical routines used to generate the figures and the finite-graph spectra can be provided by the author upon reasonable request.
	
	\begin{acknowledgments}
		The author acknowledges support from Conselho Nacional de Desenvolvimento Cient\'{i}fico e Tecnol\'{o}gico (CNPq) (grants 306308/2022-3), Funda\c c\~ao de Amparo \`a Pesquisa e ao Desenvolvimento Cient\'{i}fico e Tecnol\'{o}gico do Maranh\~ao (FAPEMA) (grants UNIVERSAL-06395/22), and Coordena\c c\~ao de Aperfei\c coamento de Pessoal de N\'{i}vel Superior (CAPES) - Brazil (Finance Code 001).
	\end{acknowledgments}


\begin{thebibliography}{29}%
	\makeatletter
	\providecommand \@ifxundefined [1]{%
		\@ifx{#1\undefined}
	}%
	\providecommand \@ifnum [1]{%
		\ifnum #1\expandafter \@firstoftwo
		\else \expandafter \@secondoftwo
		\fi
	}%
	\providecommand \@ifx [1]{%
		\ifx #1\expandafter \@firstoftwo
		\else \expandafter \@secondoftwo
		\fi
	}%
	\providecommand \natexlab [1]{#1}%
	\providecommand \enquote  [1]{``#1''}%
	\providecommand \bibnamefont  [1]{#1}%
	\providecommand \bibfnamefont [1]{#1}%
	\providecommand \citenamefont [1]{#1}%
	\providecommand \href@noop [0]{\@secondoftwo}%
	\providecommand \href [0]{\begingroup \@sanitize@url \@href}%
	\providecommand \@href[1]{\@@startlink{#1}\@@href}%
	\providecommand \@@href[1]{\endgroup#1\@@endlink}%
	\providecommand \@sanitize@url [0]{\catcode `\\12\catcode `\$12\catcode
		`\&12\catcode `\#12\catcode `\^12\catcode `\_12\catcode `\%12\relax}%
	\providecommand \@@startlink[1]{}%
	\providecommand \@@endlink[0]{}%
	\providecommand \url  [0]{\begingroup\@sanitize@url \@url }%
	\providecommand \@url [1]{\endgroup\@href {#1}{\urlprefix }}%
	\providecommand \urlprefix  [0]{URL }%
	\providecommand \Eprint [0]{\href }%
	\providecommand \doibase [0]{https://doi.org/}%
	\providecommand \selectlanguage [0]{\@gobble}%
	\providecommand \bibinfo  [0]{\@secondoftwo}%
	\providecommand \bibfield  [0]{\@secondoftwo}%
	\providecommand \translation [1]{[#1]}%
	\providecommand \BibitemOpen [0]{}%
	\providecommand \bibitemStop [0]{}%
	\providecommand \bibitemNoStop [0]{.\EOS\space}%
	\providecommand \EOS [0]{\spacefactor3000\relax}%
	\providecommand \BibitemShut  [1]{\csname bibitem#1\endcsname}%
	\let\auto@bib@innerbib\@empty
	\bibitem [{\citenamefont {Katanaev}\ and\ \citenamefont
		{Volovich}(1992)}]{Katanaev1992}%
	\BibitemOpen
	\bibfield  {author} {\bibinfo {author} {\bibfnamefont {M.}~\bibnamefont
			{Katanaev}}\ and\ \bibinfo {author} {\bibfnamefont {I.}~\bibnamefont
			{Volovich}},\ }\href
	{https://doi.org/https://doi.org/10.1016/0003-4916(52)90040-7} {\bibfield
		{journal} {\bibinfo  {journal} {Annals of Physics}\ }\textbf {\bibinfo
			{volume} {216}},\ \bibinfo {pages} {1} (\bibinfo {year} {1992})}\BibitemShut
	{NoStop}%
	\bibitem [{\citenamefont {Celi}\ \emph {et~al.}(2014)\citenamefont {Celi},
		\citenamefont {Massignan}, \citenamefont {Ruseckas}, \citenamefont {Goldman},
		\citenamefont {Spielman}, \citenamefont {Juzeli\=unas},\ and\ \citenamefont
		{Lewenstein}}]{Celi2014}%
	\BibitemOpen
	\bibfield  {author} {\bibinfo {author} {\bibfnamefont {A.}~\bibnamefont
			{Celi}}, \bibinfo {author} {\bibfnamefont {P.}~\bibnamefont {Massignan}},
		\bibinfo {author} {\bibfnamefont {J.}~\bibnamefont {Ruseckas}}, \bibinfo
		{author} {\bibfnamefont {N.}~\bibnamefont {Goldman}}, \bibinfo {author}
		{\bibfnamefont {I.~B.}\ \bibnamefont {Spielman}}, \bibinfo {author}
		{\bibfnamefont {G.}~\bibnamefont {Juzeli\=unas}},\ and\ \bibinfo {author}
		{\bibfnamefont {M.}~\bibnamefont {Lewenstein}},\ }\href
	{https://doi.org/10.1103/PhysRevLett.112.043001} {\bibfield  {journal}
		{\bibinfo  {journal} {Phys. Rev. Lett.}\ }\textbf {\bibinfo {volume} {112}},\
		\bibinfo {pages} {043001} (\bibinfo {year} {2014})}\BibitemShut {NoStop}%
	\bibitem [{\citenamefont {Yuan}\ \emph {et~al.}(2018)\citenamefont {Yuan},
		\citenamefont {Lin}, \citenamefont {Xiao},\ and\ \citenamefont
		{Fan}}]{Yuan2018}%
	\BibitemOpen
	\bibfield  {author} {\bibinfo {author} {\bibfnamefont {L.}~\bibnamefont
			{Yuan}}, \bibinfo {author} {\bibfnamefont {Q.}~\bibnamefont {Lin}}, \bibinfo
		{author} {\bibfnamefont {M.}~\bibnamefont {Xiao}},\ and\ \bibinfo {author}
		{\bibfnamefont {S.}~\bibnamefont {Fan}},\ }\href
	{https://doi.org/10.1364/OPTICA.5.001396} {\bibfield  {journal} {\bibinfo
			{journal} {Optica}\ }\textbf {\bibinfo {volume} {5}},\ \bibinfo {pages}
		{1396} (\bibinfo {year} {2018})}\BibitemShut {NoStop}%
	\bibitem [{\citenamefont {Ozawa}\ and\ \citenamefont
		{Price}(2019)}]{OzawaPrice2019}%
	\BibitemOpen
	\bibfield  {author} {\bibinfo {author} {\bibfnamefont {T.}~\bibnamefont
			{Ozawa}}\ and\ \bibinfo {author} {\bibfnamefont {H.~M.}\ \bibnamefont
			{Price}},\ }\href {https://doi.org/10.1038/s42254-019-0045-3} {\bibfield
		{journal} {\bibinfo  {journal} {Nat. Rev. Phys.}\ }\textbf {\bibinfo {volume}
			{1}},\ \bibinfo {pages} {349} (\bibinfo {year} {2019})}\BibitemShut {NoStop}%
	\bibitem [{\citenamefont {Ozawa}\ \emph {et~al.}(2019)\citenamefont {Ozawa},
		\citenamefont {Price}, \citenamefont {Amo}, \citenamefont {Goldman},
		\citenamefont {Hafezi}, \citenamefont {Lu}, \citenamefont {Rechtsman},
		\citenamefont {Schuster}, \citenamefont {Simon}, \citenamefont {Zilberberg},\
		and\ \citenamefont {Carusotto}}]{Ozawa2019}%
	\BibitemOpen
	\bibfield  {author} {\bibinfo {author} {\bibfnamefont {T.}~\bibnamefont
			{Ozawa}}, \bibinfo {author} {\bibfnamefont {H.~M.}\ \bibnamefont {Price}},
		\bibinfo {author} {\bibfnamefont {A.}~\bibnamefont {Amo}}, \bibinfo {author}
		{\bibfnamefont {N.}~\bibnamefont {Goldman}}, \bibinfo {author} {\bibfnamefont
			{M.}~\bibnamefont {Hafezi}}, \bibinfo {author} {\bibfnamefont
			{L.}~\bibnamefont {Lu}}, \bibinfo {author} {\bibfnamefont {M.~C.}\
			\bibnamefont {Rechtsman}}, \bibinfo {author} {\bibfnamefont {D.}~\bibnamefont
			{Schuster}}, \bibinfo {author} {\bibfnamefont {J.}~\bibnamefont {Simon}},
		\bibinfo {author} {\bibfnamefont {O.}~\bibnamefont {Zilberberg}},\ and\
		\bibinfo {author} {\bibfnamefont {I.}~\bibnamefont {Carusotto}},\ }\href
	{https://doi.org/10.1103/RevModPhys.91.015006} {\bibfield  {journal}
		{\bibinfo  {journal} {Rev. Mod. Phys.}\ }\textbf {\bibinfo {volume} {91}},\
		\bibinfo {pages} {015006} (\bibinfo {year} {2019})}\BibitemShut {NoStop}%
	\bibitem [{\citenamefont {Lustig}\ \emph {et~al.}(2019)\citenamefont {Lustig},
		\citenamefont {Weimann}, \citenamefont {Plotnik}, \citenamefont {Lumer},
		\citenamefont {Bandres}, \citenamefont {Szameit},\ and\ \citenamefont
		{Segev}}]{Lustig2019}%
	\BibitemOpen
	\bibfield  {author} {\bibinfo {author} {\bibfnamefont {E.}~\bibnamefont
			{Lustig}}, \bibinfo {author} {\bibfnamefont {S.}~\bibnamefont {Weimann}},
		\bibinfo {author} {\bibfnamefont {Y.}~\bibnamefont {Plotnik}}, \bibinfo
		{author} {\bibfnamefont {Y.}~\bibnamefont {Lumer}}, \bibinfo {author}
		{\bibfnamefont {M.~A.}\ \bibnamefont {Bandres}}, \bibinfo {author}
		{\bibfnamefont {A.}~\bibnamefont {Szameit}},\ and\ \bibinfo {author}
		{\bibfnamefont {M.}~\bibnamefont {Segev}},\ }\href
	{https://doi.org/10.1038/s41586-019-0943-7} {\bibfield  {journal} {\bibinfo
			{journal} {Nature}\ }\textbf {\bibinfo {volume} {567}},\ \bibinfo {pages}
		{356} (\bibinfo {year} {2019})}\BibitemShut {NoStop}%
	\bibitem [{\citenamefont {Arg\"uello-Luengo}\ \emph {et~al.}(2024)\citenamefont
		{Arg\"uello-Luengo}, \citenamefont {Bhattacharya}, \citenamefont {Celi},
		\citenamefont {Chhajlany}, \citenamefont {Grass}, \citenamefont
		{P\l{}odzie\'n}, \citenamefont {Rakshit}, \citenamefont {Salamon},
		\citenamefont {Stornati}, \citenamefont {Tarruell}, \citenamefont
		{Lewenstein} \emph {et~al.}}]{ArguelloLuengo2024}%
	\BibitemOpen
	\bibfield  {author} {\bibinfo {author} {\bibfnamefont {J.}~\bibnamefont
			{Arg\"uello-Luengo}}, \bibinfo {author} {\bibfnamefont {U.}~\bibnamefont
			{Bhattacharya}}, \bibinfo {author} {\bibfnamefont {A.}~\bibnamefont {Celi}},
		\bibinfo {author} {\bibfnamefont {R.~W.}\ \bibnamefont {Chhajlany}}, \bibinfo
		{author} {\bibfnamefont {T.}~\bibnamefont {Grass}}, \bibinfo {author}
		{\bibfnamefont {M.}~\bibnamefont {P\l{}odzie\'n}}, \bibinfo {author}
		{\bibfnamefont {D.}~\bibnamefont {Rakshit}}, \bibinfo {author} {\bibfnamefont
			{T.}~\bibnamefont {Salamon}}, \bibinfo {author} {\bibfnamefont
			{P.}~\bibnamefont {Stornati}}, \bibinfo {author} {\bibfnamefont
			{L.}~\bibnamefont {Tarruell}}, \bibinfo {author} {\bibfnamefont
			{M.}~\bibnamefont {Lewenstein}}, \emph {et~al.},\ }\href
	{https://doi.org/10.1038/s42005-024-01636-3} {\bibfield  {journal} {\bibinfo
			{journal} {Commun. Phys.}\ }\textbf {\bibinfo {volume} {7}},\ \bibinfo
		{pages} {143} (\bibinfo {year} {2024})}\BibitemShut {NoStop}%
	\bibitem [{\citenamefont {Ningyuan}\ \emph {et~al.}(2015)\citenamefont
		{Ningyuan}, \citenamefont {Owens}, \citenamefont {Sommer}, \citenamefont
		{Schuster},\ and\ \citenamefont {Simon}}]{Ningyuan2015}%
	\BibitemOpen
	\bibfield  {author} {\bibinfo {author} {\bibfnamefont {J.}~\bibnamefont
			{Ningyuan}}, \bibinfo {author} {\bibfnamefont {C.}~\bibnamefont {Owens}},
		\bibinfo {author} {\bibfnamefont {A.}~\bibnamefont {Sommer}}, \bibinfo
		{author} {\bibfnamefont {D.}~\bibnamefont {Schuster}},\ and\ \bibinfo
		{author} {\bibfnamefont {J.}~\bibnamefont {Simon}},\ }\href
	{https://doi.org/10.1103/PhysRevX.5.021031} {\bibfield  {journal} {\bibinfo
			{journal} {Phys. Rev. X}\ }\textbf {\bibinfo {volume} {5}},\ \bibinfo {pages}
		{021031} (\bibinfo {year} {2015})}\BibitemShut {NoStop}%
	\bibitem [{\citenamefont {Lee}\ \emph {et~al.}(2018)\citenamefont {Lee},
		\citenamefont {Imhof}, \citenamefont {Berger}, \citenamefont {Bayer},
		\citenamefont {Brehm}, \citenamefont {Molenkamp}, \citenamefont {Kiessling},\
		and\ \citenamefont {Thomale}}]{Lee2018}%
	\BibitemOpen
	\bibfield  {author} {\bibinfo {author} {\bibfnamefont {C.~H.}\ \bibnamefont
			{Lee}}, \bibinfo {author} {\bibfnamefont {S.}~\bibnamefont {Imhof}}, \bibinfo
		{author} {\bibfnamefont {C.}~\bibnamefont {Berger}}, \bibinfo {author}
		{\bibfnamefont {F.}~\bibnamefont {Bayer}}, \bibinfo {author} {\bibfnamefont
			{J.}~\bibnamefont {Brehm}}, \bibinfo {author} {\bibfnamefont {L.~W.}\
			\bibnamefont {Molenkamp}}, \bibinfo {author} {\bibfnamefont {T.}~\bibnamefont
			{Kiessling}},\ and\ \bibinfo {author} {\bibfnamefont {R.}~\bibnamefont
			{Thomale}},\ }\href {https://doi.org/10.1038/s42005-018-0035-2} {\bibfield
		{journal} {\bibinfo  {journal} {Commun. Phys.}\ }\textbf {\bibinfo {volume}
			{1}},\ \bibinfo {pages} {39} (\bibinfo {year} {2018})}\BibitemShut {NoStop}%
	\bibitem [{\citenamefont {Koll\'{a}r}\ \emph {et~al.}(2019)\citenamefont
		{Koll\'{a}r}, \citenamefont {Fitzpatrick},\ and\ \citenamefont
		{Houck}}]{Kollar2019}%
	\BibitemOpen
	\bibfield  {author} {\bibinfo {author} {\bibfnamefont {A.~J.}\ \bibnamefont
			{Koll\'{a}r}}, \bibinfo {author} {\bibfnamefont {M.}~\bibnamefont
			{Fitzpatrick}},\ and\ \bibinfo {author} {\bibfnamefont {A.~A.}\ \bibnamefont
			{Houck}},\ }\href {https://doi.org/10.1038/s41586-019-1348-3} {\bibfield
		{journal} {\bibinfo  {journal} {Nature}\ }\textbf {\bibinfo {volume} {571}},\
		\bibinfo {pages} {45} (\bibinfo {year} {2019})}\BibitemShut {NoStop}%
	\bibitem [{\citenamefont {Boettcher}\ \emph {et~al.}(2020)\citenamefont
		{Boettcher}, \citenamefont {Bienias}, \citenamefont {Belyansky},
		\citenamefont {Koll\'{a}r},\ and\ \citenamefont {Gorshkov}}]{Boettcher2020}%
	\BibitemOpen
	\bibfield  {author} {\bibinfo {author} {\bibfnamefont {I.}~\bibnamefont
			{Boettcher}}, \bibinfo {author} {\bibfnamefont {P.}~\bibnamefont {Bienias}},
		\bibinfo {author} {\bibfnamefont {R.}~\bibnamefont {Belyansky}}, \bibinfo
		{author} {\bibfnamefont {A.~J.}\ \bibnamefont {Koll\'{a}r}},\ and\ \bibinfo
		{author} {\bibfnamefont {A.~V.}\ \bibnamefont {Gorshkov}},\ }\href
	{https://doi.org/10.1103/PhysRevA.102.032208} {\bibfield  {journal} {\bibinfo
			{journal} {Phys. Rev. A}\ }\textbf {\bibinfo {volume} {102}},\ \bibinfo
		{pages} {032208} (\bibinfo {year} {2020})}\BibitemShut {NoStop}%
	\bibitem [{\citenamefont {Wallraff}\ \emph {et~al.}(2004)\citenamefont
		{Wallraff}, \citenamefont {Schuster}, \citenamefont {Blais}, \citenamefont
		{Frunzio}, \citenamefont {Huang}, \citenamefont {Majer}, \citenamefont
		{Kumar}, \citenamefont {Girvin},\ and\ \citenamefont
		{Schoelkopf}}]{Wallraff2004}%
	\BibitemOpen
	\bibfield  {author} {\bibinfo {author} {\bibfnamefont {A.}~\bibnamefont
			{Wallraff}}, \bibinfo {author} {\bibfnamefont {D.~I.}\ \bibnamefont
			{Schuster}}, \bibinfo {author} {\bibfnamefont {A.}~\bibnamefont {Blais}},
		\bibinfo {author} {\bibfnamefont {L.}~\bibnamefont {Frunzio}}, \bibinfo
		{author} {\bibfnamefont {R.-S.}\ \bibnamefont {Huang}}, \bibinfo {author}
		{\bibfnamefont {J.}~\bibnamefont {Majer}}, \bibinfo {author} {\bibfnamefont
			{S.}~\bibnamefont {Kumar}}, \bibinfo {author} {\bibfnamefont {S.~M.}\
			\bibnamefont {Girvin}},\ and\ \bibinfo {author} {\bibfnamefont {R.~J.}\
			\bibnamefont {Schoelkopf}},\ }\href {https://doi.org/10.1038/nature02851}
	{\bibfield  {journal} {\bibinfo  {journal} {Nature}\ }\textbf {\bibinfo
			{volume} {431}},\ \bibinfo {pages} {162} (\bibinfo {year}
		{2004})}\BibitemShut {NoStop}%
	\bibitem [{\citenamefont {Nigg}\ \emph {et~al.}(2012)\citenamefont {Nigg},
		\citenamefont {Paik}, \citenamefont {Vlastakis}, \citenamefont {Kirchmair},
		\citenamefont {Shankar}, \citenamefont {Frunzio}, \citenamefont {Devoret},
		\citenamefont {Schoelkopf},\ and\ \citenamefont {Girvin}}]{Nigg2012}%
	\BibitemOpen
	\bibfield  {author} {\bibinfo {author} {\bibfnamefont {S.~E.}\ \bibnamefont
			{Nigg}}, \bibinfo {author} {\bibfnamefont {H.}~\bibnamefont {Paik}}, \bibinfo
		{author} {\bibfnamefont {B.}~\bibnamefont {Vlastakis}}, \bibinfo {author}
		{\bibfnamefont {G.}~\bibnamefont {Kirchmair}}, \bibinfo {author}
		{\bibfnamefont {S.}~\bibnamefont {Shankar}}, \bibinfo {author} {\bibfnamefont
			{L.}~\bibnamefont {Frunzio}}, \bibinfo {author} {\bibfnamefont {M.~H.}\
			\bibnamefont {Devoret}}, \bibinfo {author} {\bibfnamefont {R.~J.}\
			\bibnamefont {Schoelkopf}},\ and\ \bibinfo {author} {\bibfnamefont {S.~M.}\
			\bibnamefont {Girvin}},\ }\href
	{https://doi.org/10.1103/PhysRevLett.108.240502} {\bibfield  {journal}
		{\bibinfo  {journal} {Phys. Rev. Lett.}\ }\textbf {\bibinfo {volume} {108}},\
		\bibinfo {pages} {240502} (\bibinfo {year} {2012})}\BibitemShut {NoStop}%
	\bibitem [{\citenamefont {Vool}\ and\ \citenamefont
		{Devoret}(2017)}]{Vool2017}%
	\BibitemOpen
	\bibfield  {author} {\bibinfo {author} {\bibfnamefont {U.}~\bibnamefont
			{Vool}}\ and\ \bibinfo {author} {\bibfnamefont {M.}~\bibnamefont {Devoret}},\
	}\href {https://doi.org/10.1002/cta.2359} {\bibfield  {journal} {\bibinfo
			{journal} {Int. J. Circuit Theory Appl.}\ }\textbf {\bibinfo {volume} {45}},\
		\bibinfo {pages} {897} (\bibinfo {year} {2017})}\BibitemShut {NoStop}%
	\bibitem [{\citenamefont {Koch}\ \emph {et~al.}(2007)\citenamefont {Koch},
		\citenamefont {Yu}, \citenamefont {Gambetta}, \citenamefont {Houck},
		\citenamefont {Schuster}, \citenamefont {Majer}, \citenamefont {Blais},
		\citenamefont {Devoret}, \citenamefont {Girvin},\ and\ \citenamefont
		{Schoelkopf}}]{Koch2007}%
	\BibitemOpen
	\bibfield  {author} {\bibinfo {author} {\bibfnamefont {J.}~\bibnamefont
			{Koch}}, \bibinfo {author} {\bibfnamefont {T.~M.}\ \bibnamefont {Yu}},
		\bibinfo {author} {\bibfnamefont {J.}~\bibnamefont {Gambetta}}, \bibinfo
		{author} {\bibfnamefont {A.~A.}\ \bibnamefont {Houck}}, \bibinfo {author}
		{\bibfnamefont {D.~I.}\ \bibnamefont {Schuster}}, \bibinfo {author}
		{\bibfnamefont {J.}~\bibnamefont {Majer}}, \bibinfo {author} {\bibfnamefont
			{A.}~\bibnamefont {Blais}}, \bibinfo {author} {\bibfnamefont {M.~H.}\
			\bibnamefont {Devoret}}, \bibinfo {author} {\bibfnamefont {S.~M.}\
			\bibnamefont {Girvin}},\ and\ \bibinfo {author} {\bibfnamefont {R.~J.}\
			\bibnamefont {Schoelkopf}},\ }\href
	{https://doi.org/10.1103/PhysRevA.76.042319} {\bibfield  {journal} {\bibinfo
			{journal} {Phys. Rev. A}\ }\textbf {\bibinfo {volume} {76}},\ \bibinfo
		{pages} {042319} (\bibinfo {year} {2007})}\BibitemShut {NoStop}%
	\bibitem [{\citenamefont {Houck}\ \emph {et~al.}(2012)\citenamefont {Houck},
		\citenamefont {T\"ureci},\ and\ \citenamefont {Koch}}]{Houck2012}%
	\BibitemOpen
	\bibfield  {author} {\bibinfo {author} {\bibfnamefont {A.~A.}\ \bibnamefont
			{Houck}}, \bibinfo {author} {\bibfnamefont {H.~E.}\ \bibnamefont
			{T\"ureci}},\ and\ \bibinfo {author} {\bibfnamefont {J.}~\bibnamefont
			{Koch}},\ }\href {https://doi.org/10.1038/nphys2251} {\bibfield  {journal}
		{\bibinfo  {journal} {Nat. Phys.}\ }\textbf {\bibinfo {volume} {8}},\
		\bibinfo {pages} {292} (\bibinfo {year} {2012})}\BibitemShut {NoStop}%
	\bibitem [{\citenamefont {Blais}\ \emph {et~al.}(2021)\citenamefont {Blais},
		\citenamefont {Grimsmo}, \citenamefont {Girvin},\ and\ \citenamefont
		{Wallraff}}]{Blais2021}%
	\BibitemOpen
	\bibfield  {author} {\bibinfo {author} {\bibfnamefont {A.}~\bibnamefont
			{Blais}}, \bibinfo {author} {\bibfnamefont {A.~L.}\ \bibnamefont {Grimsmo}},
		\bibinfo {author} {\bibfnamefont {S.~M.}\ \bibnamefont {Girvin}},\ and\
		\bibinfo {author} {\bibfnamefont {A.}~\bibnamefont {Wallraff}},\ }\href
	{https://doi.org/10.1103/RevModPhys.93.025005} {\bibfield  {journal}
		{\bibinfo  {journal} {Rev. Mod. Phys.}\ }\textbf {\bibinfo {volume} {93}},\
		\bibinfo {pages} {025005} (\bibinfo {year} {2021})}\BibitemShut {NoStop}%
	\bibitem [{\citenamefont {Silva~Netto}\ and\ \citenamefont
		{Furtado}(2008)}]{SilvaNettoFurtado2008}%
	\BibitemOpen
	\bibfield  {author} {\bibinfo {author} {\bibfnamefont {A.~L.}\ \bibnamefont
			{Silva~Netto}}\ and\ \bibinfo {author} {\bibfnamefont {C.}~\bibnamefont
			{Furtado}},\ }\href {https://doi.org/10.1088/0953-8984/20/12/125209}
	{\bibfield  {journal} {\bibinfo  {journal} {J. Phys.: Condens. Matter}\
		}\textbf {\bibinfo {volume} {20}},\ \bibinfo {pages} {125209} (\bibinfo
		{year} {2008})}\BibitemShut {NoStop}%
	\bibitem [{\citenamefont {Bakke}\ and\ \citenamefont
		{Moraes}(2012)}]{BakkeMoraes2012PLA}%
	\BibitemOpen
	\bibfield  {author} {\bibinfo {author} {\bibfnamefont {K.}~\bibnamefont
			{Bakke}}\ and\ \bibinfo {author} {\bibfnamefont {F.}~\bibnamefont {Moraes}},\
	}\href@noop {} {\bibfield  {journal} {\bibinfo  {journal} {Phys. Lett. A}\
		}\textbf {\bibinfo {volume} {376}},\ \bibinfo {pages} {2838} (\bibinfo {year}
		{2012})}\BibitemShut {NoStop}%
	\bibitem [{\citenamefont {Silva}(2026{\natexlab{a}})}]{Silva2026AnnPhys}%
	\BibitemOpen
	\bibfield  {author} {\bibinfo {author} {\bibfnamefont {E.~O.}\ \bibnamefont
			{Silva}},\ }\href {https://doi.org/10.1002/andp.202500593} {\bibfield
		{journal} {\bibinfo  {journal} {Ann. Phys. (Berlin)}\ }\textbf {\bibinfo
			{volume} {538}},\ \bibinfo {pages} {e00593} (\bibinfo {year}
		{2026}{\natexlab{a}})}\BibitemShut {NoStop}%
	\bibitem [{\citenamefont {Silva}(2026{\natexlab{b}})}]{Silva2026OQE}%
	\BibitemOpen
	\bibfield  {author} {\bibinfo {author} {\bibfnamefont {E.~O.}\ \bibnamefont
			{Silva}},\ }\href {https://doi.org/10.1007/s11082-026-08693-8} {\bibfield
		{journal} {\bibinfo  {journal} {Opt. Quantum Electron.}\ }\textbf {\bibinfo
			{volume} {58}},\ \bibinfo {pages} {113} (\bibinfo {year}
		{2026}{\natexlab{b}})}\BibitemShut {NoStop}%
	\bibitem [{\citenamefont {Koch}\ \emph {et~al.}(2010)\citenamefont {Koch},
		\citenamefont {Houck}, \citenamefont {Le~Hur},\ and\ \citenamefont
		{Girvin}}]{Koch2010}%
	\BibitemOpen
	\bibfield  {author} {\bibinfo {author} {\bibfnamefont {J.}~\bibnamefont
			{Koch}}, \bibinfo {author} {\bibfnamefont {A.~A.}\ \bibnamefont {Houck}},
		\bibinfo {author} {\bibfnamefont {K.}~\bibnamefont {Le~Hur}},\ and\ \bibinfo
		{author} {\bibfnamefont {S.~M.}\ \bibnamefont {Girvin}},\ }\href
	{https://doi.org/10.1103/PhysRevA.82.043811} {\bibfield  {journal} {\bibinfo
			{journal} {Phys. Rev. A}\ }\textbf {\bibinfo {volume} {82}},\ \bibinfo
		{pages} {043811} (\bibinfo {year} {2010})}\BibitemShut {NoStop}%
	\bibitem [{\citenamefont {Roushan}\ \emph {et~al.}(2017)\citenamefont
		{Roushan}, \citenamefont {Neill}, \citenamefont {Megrant}, \citenamefont
		{Chen}, \citenamefont {Babbush}, \citenamefont {Barends}, \citenamefont
		{Campbell}, \citenamefont {Chen}, \citenamefont {Chiaro}, \citenamefont
		{Dunsworth}, \citenamefont {Fowler}, \citenamefont {Jeffrey}, \citenamefont
		{Kelly}, \citenamefont {Lucero}, \citenamefont {Mutus}, \citenamefont
		{O'Malley}, \citenamefont {Neeley}, \citenamefont {Quintana}, \citenamefont
		{Sank}, \citenamefont {Vainsencher}, \citenamefont {Wenner}, \citenamefont
		{White}, \citenamefont {Kapit}, \citenamefont {Neven},\ and\ \citenamefont
		{Martinis}}]{Roushan2017}%
	\BibitemOpen
	\bibfield  {author} {\bibinfo {author} {\bibfnamefont {P.}~\bibnamefont
			{Roushan}}, \bibinfo {author} {\bibfnamefont {C.}~\bibnamefont {Neill}},
		\bibinfo {author} {\bibfnamefont {A.}~\bibnamefont {Megrant}}, \bibinfo
		{author} {\bibfnamefont {Y.}~\bibnamefont {Chen}}, \bibinfo {author}
		{\bibfnamefont {R.}~\bibnamefont {Babbush}}, \bibinfo {author} {\bibfnamefont
			{R.}~\bibnamefont {Barends}}, \bibinfo {author} {\bibfnamefont
			{B.}~\bibnamefont {Campbell}}, \bibinfo {author} {\bibfnamefont
			{Z.}~\bibnamefont {Chen}}, \bibinfo {author} {\bibfnamefont {B.}~\bibnamefont
			{Chiaro}}, \bibinfo {author} {\bibfnamefont {A.}~\bibnamefont {Dunsworth}},
		\bibinfo {author} {\bibfnamefont {A.}~\bibnamefont {Fowler}}, \bibinfo
		{author} {\bibfnamefont {E.}~\bibnamefont {Jeffrey}}, \bibinfo {author}
		{\bibfnamefont {J.}~\bibnamefont {Kelly}}, \bibinfo {author} {\bibfnamefont
			{E.}~\bibnamefont {Lucero}}, \bibinfo {author} {\bibfnamefont
			{J.}~\bibnamefont {Mutus}}, \bibinfo {author} {\bibfnamefont {P.~J.~J.}\
			\bibnamefont {O'Malley}}, \bibinfo {author} {\bibfnamefont {M.}~\bibnamefont
			{Neeley}}, \bibinfo {author} {\bibfnamefont {C.}~\bibnamefont {Quintana}},
		\bibinfo {author} {\bibfnamefont {D.}~\bibnamefont {Sank}}, \bibinfo {author}
		{\bibfnamefont {A.}~\bibnamefont {Vainsencher}}, \bibinfo {author}
		{\bibfnamefont {J.}~\bibnamefont {Wenner}}, \bibinfo {author} {\bibfnamefont
			{T.}~\bibnamefont {White}}, \bibinfo {author} {\bibfnamefont
			{E.}~\bibnamefont {Kapit}}, \bibinfo {author} {\bibfnamefont
			{H.}~\bibnamefont {Neven}},\ and\ \bibinfo {author} {\bibfnamefont
			{J.}~\bibnamefont {Martinis}},\ }\href {https://doi.org/10.1038/nphys3930}
	{\bibfield  {journal} {\bibinfo  {journal} {Nat. Phys.}\ }\textbf {\bibinfo
			{volume} {13}},\ \bibinfo {pages} {146} (\bibinfo {year} {2017})}\BibitemShut
	{NoStop}%
	\bibitem [{\citenamefont {Kirchmair}\ \emph {et~al.}(2013)\citenamefont
		{Kirchmair}, \citenamefont {Vlastakis}, \citenamefont {Leghtas},
		\citenamefont {Nigg}, \citenamefont {Paik}, \citenamefont {Ginossar},
		\citenamefont {Mirrahimi}, \citenamefont {Frunzio}, \citenamefont {Girvin},\
		and\ \citenamefont {Schoelkopf}}]{Kirchmair2013}%
	\BibitemOpen
	\bibfield  {author} {\bibinfo {author} {\bibfnamefont {G.}~\bibnamefont
			{Kirchmair}}, \bibinfo {author} {\bibfnamefont {B.}~\bibnamefont
			{Vlastakis}}, \bibinfo {author} {\bibfnamefont {Z.}~\bibnamefont {Leghtas}},
		\bibinfo {author} {\bibfnamefont {S.~E.}\ \bibnamefont {Nigg}}, \bibinfo
		{author} {\bibfnamefont {H.}~\bibnamefont {Paik}}, \bibinfo {author}
		{\bibfnamefont {E.}~\bibnamefont {Ginossar}}, \bibinfo {author}
		{\bibfnamefont {M.}~\bibnamefont {Mirrahimi}}, \bibinfo {author}
		{\bibfnamefont {L.}~\bibnamefont {Frunzio}}, \bibinfo {author} {\bibfnamefont
			{S.~M.}\ \bibnamefont {Girvin}},\ and\ \bibinfo {author} {\bibfnamefont
			{R.~J.}\ \bibnamefont {Schoelkopf}},\ }\href
	{https://doi.org/10.1038/nature11902} {\bibfield  {journal} {\bibinfo
			{journal} {Nature}\ }\textbf {\bibinfo {volume} {495}},\ \bibinfo {pages}
		{205} (\bibinfo {year} {2013})}\BibitemShut {NoStop}%
	\bibitem [{\citenamefont {Song}\ \emph {et~al.}(2024)\citenamefont {Song},
		\citenamefont {Ruan}, \citenamefont {Ding}, \citenamefont {Li}, \citenamefont
		{Chen}, \citenamefont {Huang}, \citenamefont {Kuang}, \citenamefont {Zhao},
		\citenamefont {Tsai}, \citenamefont {Jing}, \citenamefont {Yang},
		\citenamefont {Nori}, \citenamefont {Zheng}, \citenamefont {Liu},
		\citenamefont {Zhang},\ and\ \citenamefont {Peng}}]{Song2024}%
	\BibitemOpen
	\bibfield  {author} {\bibinfo {author} {\bibfnamefont {P.}~\bibnamefont
			{Song}}, \bibinfo {author} {\bibfnamefont {X.}~\bibnamefont {Ruan}}, \bibinfo
		{author} {\bibfnamefont {H.}~\bibnamefont {Ding}}, \bibinfo {author}
		{\bibfnamefont {S.}~\bibnamefont {Li}}, \bibinfo {author} {\bibfnamefont
			{M.}~\bibnamefont {Chen}}, \bibinfo {author} {\bibfnamefont {R.}~\bibnamefont
			{Huang}}, \bibinfo {author} {\bibfnamefont {L.-M.}\ \bibnamefont {Kuang}},
		\bibinfo {author} {\bibfnamefont {Q.}~\bibnamefont {Zhao}}, \bibinfo {author}
		{\bibfnamefont {J.-S.}\ \bibnamefont {Tsai}}, \bibinfo {author}
		{\bibfnamefont {H.}~\bibnamefont {Jing}}, \bibinfo {author} {\bibfnamefont
			{L.}~\bibnamefont {Yang}}, \bibinfo {author} {\bibfnamefont {F.}~\bibnamefont
			{Nori}}, \bibinfo {author} {\bibfnamefont {D.}~\bibnamefont {Zheng}},
		\bibinfo {author} {\bibfnamefont {Y.-x.}\ \bibnamefont {Liu}}, \bibinfo
		{author} {\bibfnamefont {J.}~\bibnamefont {Zhang}},\ and\ \bibinfo {author}
		{\bibfnamefont {Z.}~\bibnamefont {Peng}},\ }\href
	{https://doi.org/10.1038/s41467-024-54199-w} {\bibfield  {journal} {\bibinfo
			{journal} {Nat. Commun.}\ }\textbf {\bibinfo {volume} {15}},\ \bibinfo
		{pages} {9848} (\bibinfo {year} {2024})}\BibitemShut {NoStop}%
	\bibitem [{\citenamefont {Miri}\ and\ \citenamefont {Al\`u}(2019)}]{Miri2019}%
	\BibitemOpen
	\bibfield  {author} {\bibinfo {author} {\bibfnamefont {M.-A.}\ \bibnamefont
			{Miri}}\ and\ \bibinfo {author} {\bibfnamefont {A.}~\bibnamefont {Al\`u}},\
	}\href {https://doi.org/10.1126/science.aar7709} {\bibfield  {journal}
		{\bibinfo  {journal} {Science}\ }\textbf {\bibinfo {volume} {363}},\ \bibinfo
		{pages} {eaar7709} (\bibinfo {year} {2019})}\BibitemShut {NoStop}%
	\bibitem [{\citenamefont {Ashida}\ \emph {et~al.}(2020)\citenamefont {Ashida},
		\citenamefont {Gong},\ and\ \citenamefont {Ueda}}]{Ashida2020}%
	\BibitemOpen
	\bibfield  {author} {\bibinfo {author} {\bibfnamefont {Y.}~\bibnamefont
			{Ashida}}, \bibinfo {author} {\bibfnamefont {Z.}~\bibnamefont {Gong}},\ and\
		\bibinfo {author} {\bibfnamefont {M.}~\bibnamefont {Ueda}},\ }\href
	{https://doi.org/10.1080/00018732.2021.1876991} {\bibfield  {journal}
		{\bibinfo  {journal} {Adv. Phys.}\ }\textbf {\bibinfo {volume} {69}},\
		\bibinfo {pages} {249} (\bibinfo {year} {2020})}\BibitemShut {NoStop}%
	\bibitem [{\citenamefont {Bergholtz}\ \emph {et~al.}(2021)\citenamefont
		{Bergholtz}, \citenamefont {Budich},\ and\ \citenamefont
		{Kunst}}]{Bergholtz2021}%
	\BibitemOpen
	\bibfield  {author} {\bibinfo {author} {\bibfnamefont {E.~J.}\ \bibnamefont
			{Bergholtz}}, \bibinfo {author} {\bibfnamefont {J.~C.}\ \bibnamefont
			{Budich}},\ and\ \bibinfo {author} {\bibfnamefont {F.~K.}\ \bibnamefont
			{Kunst}},\ }\href {https://doi.org/10.1103/RevModPhys.93.015005} {\bibfield
		{journal} {\bibinfo  {journal} {Rev. Mod. Phys.}\ }\textbf {\bibinfo {volume}
			{93}},\ \bibinfo {pages} {015005} (\bibinfo {year} {2021})}\BibitemShut
	{NoStop}%
	\bibitem [{\citenamefont {Naghiloo}\ \emph {et~al.}(2019)\citenamefont
		{Naghiloo}, \citenamefont {Abbasi}, \citenamefont {Joglekar},\ and\
		\citenamefont {Murch}}]{Naghiloo2019}%
	\BibitemOpen
	\bibfield  {author} {\bibinfo {author} {\bibfnamefont {M.}~\bibnamefont
			{Naghiloo}}, \bibinfo {author} {\bibfnamefont {M.}~\bibnamefont {Abbasi}},
		\bibinfo {author} {\bibfnamefont {Y.~N.}\ \bibnamefont {Joglekar}},\ and\
		\bibinfo {author} {\bibfnamefont {K.~W.}\ \bibnamefont {Murch}},\ }\href
	{https://doi.org/10.1038/s41567-019-0652-z} {\bibfield  {journal} {\bibinfo
			{journal} {Nat. Phys.}\ }\textbf {\bibinfo {volume} {15}},\ \bibinfo {pages}
		{1232} (\bibinfo {year} {2019})}\BibitemShut {NoStop}%
\end{thebibliography}
\end{document}